\documentclass[twocolumn]{aastex631}

\usepackage{CJK}
\usepackage{bm}

\shorttitle{Long-term evolution of Supercritical Accretion into BHs}
\shortauthors{Hu et al.}

\graphicspath{{./}{figures/}}

\definecolor{HHcolor}{rgb}{0.93,0.57,0.13}
\definecolor{HHcolor2}{rgb}{0.5,0.1,0.5}

\begin{document}
\begin{CJK*}{UTF8}{gbsn} 
\title{Long-term evolution of supercritical black hole accretion with 
outflows: \\a subgrid feedback model for cosmological simulations}

\correspondingauthor{Haojie Hu(胡豪杰)}
\email{hhj\_pku@pku.edu.cn}

\author[0000-0003-3143-3995]{Haojie Hu}
\affiliation{Kavli Institute for Astronomy and Astrophysics, Peking University, 5 Yiheyuan Road,  Haidian District, Beijing, 100871, PRC}
\affiliation{Department of Astronomy, School of Physics, Peking University, 5 Yiheyuan Road,  Haidian District, Beijing, 100871, PRC}

\author[0000-0001-9840-4959]{Kohei Inayoshi}
\affiliation{Kavli Institute for Astronomy and Astrophysics, Peking University, 5 Yiheyuan Road, Haidian District, Beijing, 100871, PRC}

\author[0000-0003-3633-5403]{Zolt\'an Haiman}
\affiliation{Department of Astronomy, Columbia University, New York, NY 10027, USA}

\author[0000-0001-9185-5044]{Eliot Quataert}
\affiliation{Department of Astrophysical Sciences, Princeton University, Peyton Hall, Princeton, NJ 08544, USA}

\author[0000-0003-2309-8963]{Rolf Kuiper}
\affiliation{Zentrum f\"ur Astronomie der Universit\"at Heidelberg, Institut f\"ur Theoretische Astrophysik, Albert-Ueberle-Stra$\ss$e 2, 69120 Heidelberg, Germany}

\begin{abstract}
We study the long-term evolution of the global structure of axisymmetric accretion flows onto a black hole (BH) at rates 
substantially higher than the Eddington value ($\dot{M}_{\rm Edd}$), performing two-dimensional hydrodynamical 
simulations with and without radiative diffusion. In the high-accretion optically thick limit, where the radiation energy is efficiently trapped within the inflow,
the accretion flow becomes adiabatic and comprises of turbulent gas in the equatorial region and strong bipolar outflows. 
As a result, the mass inflow rate decreases toward the center as $\dot{M}_{\rm in}\propto r^{p}$ with $p\sim 0.5-0.7$
and a small fraction of the inflowing gas feeds the nuclear BH. 
Thus, super-Eddington accretion is sustained only when a larger amount of gas is supplied from larger radii at $\ga 100-1000~\dot{M}_{\rm Edd}$. 
The global structure of the flow settles down to a quasi-steady state in millions of the orbital timescale at the BH event horizon, 
which is $\ga 10-100$ times longer than that addressed in previous (magneto-)RHD simulation studies.
Energy transport via radiative diffusion accelerates the outflow near the poles in the inner region but does not change 
the overall properties of the accretion flow compared to the cases without diffusion. Based on our simulation results, 
we provide a mechanical feedback model for super-Eddington accreting BHs.   
This can be applied as a sub-grid model in large-scale cosmological simulations that do not sufficiently 
resolve galactic nuclei, and to the formation of the heaviest gravitational-wave sources via accretion in dense environments.
\end{abstract}


\section{Introduction} \label{sec:intro}

Observations have revealed the existence of many supermassive black holes (SMBHs) at high redshift 
($z\geq 6$) with mass as large as $10^9~M_{\odot}$ 
\citep[e.g.][]{Fan2006,Mortlock2011,wu2015ultraluminous,banados2018800,matsuoka2018subaru,matsuoka2018subarub,matsuoka2018subaruc}. 
However, the formation and evolution of these SMBHs still remains unclear 
\citep[e.g.][]{king2003black,murray2005maximum,kormendy2013coevolution}.

Several formation channels of high-redshift SMBHs have been proposed \citep[see][for recent comprehensive reviews]{Inayoshi2020,Volonteri2021}.  
One possible approach is to initiate with seed BHs with mass of $\sim 10-100~M_{\odot}$, formed after the collapse 
of the first generation stars \citep{Abel2002,Bromm2002,Yoshida2008,Hirano2014}. 
It is possible for these seeds to grow into SMBHs within $\sim 1~$Gyr if the BHs continuously accrete 
at or slightly above \citep{tanaka2009assembly,Madau2014} the Eddington rate of $\dot{M}_{\rm Edd}[\equiv L_{\rm Edd}/(0.1 c^2)]$ with a $10\%$ radiative efficiency, where $L_{\rm Edd}$ is 
the Eddington luminosity, $c$ is the speed of light.
An alternative approach is to begin with heavier BH seeds with $\sim 10^4-10^6~M_{\odot}$ via 
collapse of supermassive stars (SMSs) or runaway stellar mergers in a dense star cluster
\citep{bromm2003formation,Omukai2008,devecchi2009formation,Inayoshi2014,Hirano2017,chon2020supermassive,Li2021}. 
Under special circumstances with strong H$_2$ photo-dissociating radiation
produced from nearby galaxies, violent successive halo mergers, or
high baryon-dark matter streaming motion, massive seed BHs are expected to form efficiently
and get a head start toward the SMBH regime
\citep{Omukai2001,Dijkstra2008,tanaka2014formation, Inayoshi2018, wise2019formation}.

In either case, efficient subsequent growth of the seed BHs is required to explain the $\ga10^9\, M_{\odot}$ SMBHs at early times.
The nature of rapid mass accretion onto BHs has been extensively investigated.
Observations of ultra-luminous X-ray sources \citep{king2001ultraluminous,watarai2001slim} and 
narrow-line Seyfert-1 galaxies \citep{wang1999emergent,mineshige2000slim} have provided 
several lines of evidence that both stellar mass BHs and SMBHs are able to accrete gas at rates
exceeding the Eddington rates.

Super-Eddington accretion onto stellar-mass BHs in dense gaseous environments (e.g., accretion disks of 
active galactic nuclei; hereafter AGNs) has been further explored 
\citep[][]{Mckernan2012,Stone2017,Yang2019,Secunda2019,Tagawa2020,Tagawa2021}
since the discoveries of intermediate-mass BHs detected as gravitational-wave (GW) sources \citep[][]{Abbott2021}. 
The mass spectrum of those merging BHs shows a continuous distribution, namely a single power-law + a 
Gaussian peak or a double power-law \citep[][]{Abbott2021population}, without a discontinuity that is expected 
to appear owing to (pulsational) pair-instability supernova (PISN) explosions of massive stellar progenitors
\citep[][]{Barkat1967,Fowler1964,Heger2002,Heger2003,vink2012very,Belczynski2016,Abbott2021population}.
Indeed, several GW events such as GW190521, GW190602\_175927, and GW190519\_153544 
are inferred to have their primary BH mass in the expected PISN mass gap. 
The existence of such massive BHs would not be preferred in the isolated field-binary scenario 
due to PISNe caused by single massive stars with $M_\star \sim50-120~M_{\odot}$. 
Therefore, other formation channels through hierarchical BH mergers
\citep[][]{Tagawa2021} and/or the subsequent growth of stellar remnant BHs via super-Eddington 
accretion have been considered as a solution for the existence of massive GW sources \citep[][]{Safarzadeh2020}.

Rapid mass accretion onto BHs has been investigated by analytical \citep[e.g.][]{begelman1979can} and numerical work.
In recent decades, radiation magneto-hydrodynamical  (RMHD) simulations of accreting BHs 
have shown that the BH feeding rate can exceed the critical value as long as a dense gaseous 
torus already exists or a sufficient amount of gas supply is maintained from larger radii to 
the vicinity of the BH \citep[see also][]{ohsuga2011global,jiang2014global,skadowski2015global}. 
Indeed, they found that BHs can be fed at mildly super-Eddington accretion rates of
$\sim 2-40~\dot{M}_{\rm Edd}$ at least for short durations of $\approx 10^{4-5}~t_0$, where $t_0 (\equiv GM_{\rm BH}/c^3)$ is the light-crossing timescle at half of the Schwarzschild radius ($r_{\rm Sch}$) around a BH with a mass of $M_{\rm BH}$.
However, the gravitational energy released as radiation (and/or in outflows) heats the gas surrounding the BH
and thus can quench the gas supply from the BH's gravitational influence radius.
As a result of radiative feedback, when the density of the ambient gas is low, the BH feeding rate 
is strongly suppressed and the time-averaged rate is limited below the Eddington accretion rate 
\citep[e.g.][]{ciotti2001cooling,alvarez2009accretion,
milosavljevic2009accretiona,milosavljevic2009accretionb,park2011accretion,park2012accretion,
jeon2012first}.
This  discrepancy between the results regarding the BH feeding rates and impacts of radiative feedback is owing to inconsistent treatments of numerical simulations,
most of which have focused on the dynamics of gas and radiation at {\it either} small {\it or} large scales.
This fact emphasizes the importance of understanding the global structure of accretion flows 
\citep[see a recent work by ][where both the BH event horizon scale and Bondi scale for hot accreting gas are simultaneously resolved.]{Lalakos2022}.

Recently, the global properties of super-Eddington accretion flows have been extensively studied
with radiation hydrodynamical simulations with a sufficiently large computation domain that enables us to 
capture the multi-scale physics properly \citep[e.g.,][]{inayoshi2016hyper,Takeo2018,Toyouchi2018,Takeo2020}.
Unlike the cases mentioned above, when the BH is embedded in sufficiently dense gas, 
the emergent radiation flux is substantially reduced by photon trapping and dust absorption in the flow.
Radiative feedback does not prevent the gas inflow and instead the ionized region surrounding the BH collapses.
As a result, rapid quasi-steady mass accretion exceeding the Eddington rate ($\ga 500~\dot{M}_{\rm Edd}$) through 
a dense, optically thick disk is achieved.
The extremely high accretion rate is expected to take place in dense regions such as the centers of 
high-redshift protogalaxies and/or the interiors of AGN disks even at lower redshifts.
If such intense accretion flows are maintained down to the BH event horizon and can directly feed the BH 
without significant mass loss at larger radii, this process potentially solves the problems regarding 
(1) SMBH formation and (2) the lack of the PISN mass gap on the mass spectrum of merging binary BHs.

In this paper, we investigate the properties of accretion flows onto a BH at very high rates of 
$\dot{M}\sim 100-2,000~\dot{M}_{\rm Edd}$ with two-dimensional, axisymmetric radiation hydrodynamical simulations 
extending to $\sim 10^3\,r_{\rm Sch}$. In particular, the following two issues are addressed.
First, we study the long-term evolution of high-rate accretion flows over $t > 3\times 10^6~t_0$,
which is required to reach a {\it true} steady state of the flow; namely, one viscous timescale at
$\sim 10^3~r_{\rm Sch}$.
We note that the previous simulations that studied mildly super-Eddington accreting flows were able to reach $\la10\%$ of 
this duration due to computational limitations such as short characteristic timescales of radiation transport and requirement 
of a more accurate numerical algorithm to solve the radiation-transfer equations for those systems. On the other hand, in our 
simulations where the inflow rate is much higher and the radiation energy is efficiently trapped within the inflow, radiation 
transport is well approximated by diffusion and the timescale of radiation transport becomes longer in the dense accreting 
matter, enabling longer-term simulations.
Secondly, we consider various types of initial conditions and boundary conditions to check their effects 
on the resulting accretion flow. 

The rest of this paper is organized as follows. In \S \ref{sec:method}, we describe the methodology of our simulations.
In \S \ref{sec:results}, we discuss the simulation results and present the self-similar behavior of supercritical accretion flows.
In \S \ref{sec:subgrmodel} 
 we present a subgrid model of AGN feedback for large scale simulations 
in the presence of outflows. In \S \ref{sec:summary}, 
we summarize the main conclusions of this paper.

\section{Methodology} \label{sec:method}

\subsection{Basic equations} \label{subsubsec:equations}

We study the gas dynamics at the vicinity of a BH, performing axisymmetric two-dimensional radiation
hydrodynamical simulations with the open source code {\tt PLUTO} \citep{mignone2007pluto}.  
The mass flux for the ideal fluid is computed using the Harten-Lax-vanLeer Riemann solver \citep{Harten1983}, 
and the second-order accuracy in space and time is ensured. The diffusion terms (viscosity and radiation 
transport) are treated with the explicit time integration (see below).
The basic equations in spherical coordinates $(r,\theta,\phi)$ are the equation of continuity,
\begin{equation}
\frac{\partial \rho}{\partial t}+\nabla \cdot (\rho  \boldsymbol{v})=0,
\label{eq:continuity}
\end{equation}
and the equation of motion,
\begin{equation}
\rho \frac{\partial \boldsymbol{v}}{\partial t}+\rho  (\boldsymbol{v} \cdot \nabla)  \boldsymbol{v}
=-\nabla p - \rho \nabla \Phi + \nabla  \cdot \boldsymbol{\sigma},
\label{eq:motion}
\end{equation}
where $\rho$ is the density, $\boldsymbol{v}$ is the velocity, $p$ is the total pressure, 
and $\Phi=-GM_{\mathrm{BH}}/(r-r_{\mathrm{Sch}})$ is the gravitational potential of the BH \citep[][]{B.Paczynsky}, 
and $\boldsymbol{\sigma}$ is the viscous stress tensor.
We solve the energy equation 
\begin{equation}
\frac{\partial e}{\partial t}+\nabla \cdot\left[\left(e+p\right) \boldsymbol{v} + \boldsymbol{F}_{0}\right]= 
+ (\boldsymbol{\sigma} \cdot \nabla) \boldsymbol{v},
\label{eq:energy}
\end{equation}
where $e=\frac{1}{2} \rho |\boldsymbol{v}|^{2} + \varepsilon$ is the total energy density, $\varepsilon$ is the internal energy density,
$\boldsymbol{F}_{0}$ is the radiative flux in the comoving frame, 
and the right-hand-side represents viscous heating.

In our axisymmetric simulations without magneto-hydrodynamical (MHD) effects,
angular momentum transport in the accretion flow is calculated by imposing viscosity.
The viscous stress tensor is given by
\begin{equation}
\sigma_{ij}=\rho \nu \left[ 
\left(\frac{\partial v_j}{\partial x_i} + \frac{\partial v_i}{\partial x_j}\right)
-\frac{2}{3} (\nabla \cdot \mbox{\boldmath $v$} )\delta_{ij}
\right],
\end{equation}
where $\nu$ is the shear viscosity and the bulk viscosity is neglected.
To mimic angular momentum transport associated with MHD turbulence driven by 
the magneto-rotational instability (MRI) in a disk \citep[][]{Stone1996,Balbus1991,Balbus1998}, we assume the azimuthal 
components of the shear tensor are non-zero.
In spherical polar coordinates, the tensor components are given by
\begin{equation}
\sigma_{r\phi}=\rho \nu \frac{\partial}{\partial r}\left(\frac{v_\phi}{r}\right),
\end{equation}
\begin{equation}
\sigma_{\theta \phi}=\rho \nu \frac{\sin \theta}{r}\frac{\partial}{\partial \theta}\left(\frac{v_\phi}{\sin \theta}\right),
\end{equation}
\citep[e.g.,][]{stone1999hydrodynamical,Orazio2013}.
The strength of anomalous shear viscosity is calculated with the $\alpha$-prescription \citep{shakura1973black},
\begin{equation}
\nu = \alpha \frac{c_{\rm s}^2}{\Omega_{\rm K}}f(\theta),
\end{equation}
where $\alpha$ is the viscous parameter, $\Omega_{\rm K}\equiv (GM_{\rm BH}/r^3)^{1/2}$ is the Keplerian angular frequency, 
and $c_{\rm s} = \sqrt{\gamma p/\rho}$ is the sound speed.
In our simulations, we adopt a single value of $\alpha = 0.01$ and activate the viscous process
within the flow using the function $f(\theta)$ defined by
\begin{equation}
f(\theta)=\left\{\begin{array}{ll}1 ~~& {\rm for}~~|\theta-\frac{\pi}{2}| \leq \frac{\pi}{4}, \\ 
e^{-|\cot(\theta)|} ~~& {\rm for}~~ |\theta-\frac{\pi}{2}|>\frac{\pi}{4}.
\end{array}
\right.
\label{eq:alpha}
\end{equation}
The choice of this functions is to mimic viscosity restricted inside the dense region (or disk). Since the gas density near the 
poles is several orders of magnitude lower than that near the equator, weak viscosity in the polar region hardly affects our conclusions.

\begin{deluxetable*}{cccccc|cccc}
\tablenum{1}
\tablecaption{Initial setups and results of 2D hydrodynamical simulations\label{tab:setups}}
\tablewidth{0pt}
\tablehead{
\multicolumn{6}{c}{Simulation setup} \vline& \multicolumn{4}{c}{Results}\\
\tableline
\colhead{Simulation} & \colhead{Res.} & \colhead{Inflow density} & \colhead{Boundary} & \colhead{Radiation} & \colhead{Inflow}
\vline&\colhead{$\dot{M}_0$} & \colhead{$\dot{M}_{\mathrm{BH}}$} & \colhead{Mass loading} & \colhead{$r_{{\rm tr, eff}}$}\\
\colhead{ID} & \colhead{} & \colhead{$(6\times10^{-9})$} & \colhead{Conditions} & \colhead{Diffusion} & \colhead{$2\theta_{0}$($\pi$)}
\vline&\colhead{($\dot{M}_{\mathrm{Edd}}$)} & \colhead{($\dot{M}_{\mathrm{Edd}}$)} & \colhead{factor($\beta$)} &\colhead{($r_{\mathrm{Sch}}$)}
}
\startdata
Case-H 	& 	256$\times$256 	& 10 	& High  		& NO 	& 1/2     	& 1408 	&  88 	& 15.0  	& 50\\
Case-L 	& 	256$\times$256 	& 1 	& Low	  	& NO 	& 1/2     	& 185 	&  10   	& 17.5  	& 40\\
Case-N 	& 	256$\times$256 	& 1 	& Narrow  	& NO 	& 1/6    	& 44  	&  3	 	& 13.7	& 30\\
Case-I 	& 	256$\times$256 	& 1 	& Injection  	& NO 	& 1/6    	& 320 	&  12 	& 25.7	& 45\\
Case-H-LR& 	128$\times$128 	& 10 	& High  		& NO 	& 1/2    	& 1050 	&  31 	& 32.9	& 50\\
Case-D	& 	128$\times$128 	& 10 	& High	    	& YES 	& 1/2   	& 1215 	&  39 	& 30.1	& 40\\
\enddata
\tablecomments{This table shows the initial setups and some results for 2D simulations. Density is in g$\,$cm$^{-3}$. 
$\dot{M}_0$ is the inflow rate at the outer boundary. $\dot{M}_{\mathrm{BH}}$ is the net accretion onto the BH at 
the inner boundary. $r_{\mathrm{tr},{\rm eff}}$ is the effective trapping radius inferred from the rest-frame luminosity. Here the 
`H', `L',  `N', `I', `D' represent the setup of boundary conditions and `LR' is short for low resolution. For Case-I,  a constant 
radial inflow velocity is imposed at the outer boundary in the equatorial region; otherwise 
the outer boundary conditions depend on the direction of the radial velocity (see the text).
As discussed in \S \ref{subsubsec:Diffusion}, the mass inflow and BH accretion rate show the self-similar nature and thus 
those absolute rates in the cases without radiative diffusion can be renormalized to other values as long as 
$\dot{M}_{\rm in} \gg \dot{M}_{\rm Edd}$ and $\tau \gg c/v$ hold. Note that the photon trapping radius is well-defined 
only with radiative diffusion, but we show the values for the cases without radiative diffusion for reference values.}
\end{deluxetable*}

In our simulations, we assume that the radiation pressure dominates the total pressure \citep{Begelman1978}.
This approximation is valid when the mass accretion rate significantly exceeds the Eddington rate 
and the flow is highly optically thick. The internal energy density is then given by
\begin{equation}
\varepsilon \simeq E_{\mathrm{0}}=3 P_{0},
\end{equation}
where $E_{\mathrm{0}}$ is radiation energy density, $P_{0}(\simeq p)$ is the radiation pressure,
and the adiabatic index corresponds to $\gamma \simeq 4/3$. 
We also adopt the flux-limited diffusion (FLD) approximation, where the radiation flux is calculated with
\begin{equation}
\boldsymbol{F}_{0}=-\frac{c \lambda}{\rho \kappa_{\rm es}} \nabla E_{0},
\label{eq:FLD}
\end{equation}
where $\kappa_{\rm es}$ is the opacity due to electron scattering and the flux limiter $\lambda$ is given by
\begin{equation}
\lambda=\frac{2+\mathcal{R}}{6+3 \mathcal{R}+\mathcal{R}^{2}},
\end{equation}
where $\mathcal{R}=\left|\nabla E_{0}\right| /\left(\rho \kappa_{\rm es}  E_{0}\right)$. 
In the optically thick limit ($\mathcal{R}\rightarrow 0$), the diffusion coefficient approaches $\lambda =1/3$.

The Lorentz transformation of the radiation flux between the rest frame $\boldsymbol{F}$ and 
the comoving frame $\boldsymbol{F}_{0}$ is given by 
\begin{equation}
\boldsymbol{F} = \boldsymbol{F}_0 + \boldsymbol{v}(E_{0}+P_0),
\label{eq:Lorentz_trans}
\end{equation}
Note that using this transformation, the radiation moments in the divergence term of Eq.~(\ref{eq:energy}) are replaced 
by the radiation flux $\boldsymbol{F}$ in the observer's rest-frame.
The first and second term corresponds to the diffusion and advection term, respectively.
Within dense and fast accretion flows, the two terms are comparable and the rest-frame flux decreases 
because $|\boldsymbol{F} - \boldsymbol{F}_0| \simeq \tau |(\boldsymbol{v}/c)\cdot  \boldsymbol{F}_0|$, 
where $\tau$ is the optical depth within a scale where the radiation energy density varies.
In fact, the ratio of these terms in the spherically symmetric flow (along the radial direction) is
\begin{equation}
\frac{\rm diffusion}{\rm advection} = \frac{|\boldsymbol{F}_0|}{|\boldsymbol{v}(E_{0}+P_0)|}\simeq  \frac{4\pi c r}{\kappa_{\rm es}\dot{M}} \left(\equiv \frac{r}{r_{\rm tr}}\right),
\label{eq:tr}
\end{equation}
where $\dot{M}=4\pi \rho v r^2$ is the mass flow rate.
This means that the radiative flux in the rest frame has a positive (outward) value at $r>r_{\rm tr}$
and a negative (inward) value at $r<r_{\rm tr}$, respectively.
We note that the expression of the so-called photon-trapping radius ($r_{\rm tr}\equiv \kappa_{\rm es}\dot{M}/4\pi c $) is valid for spherically symmetric flows \citep{Begelman1978}.
Nevertheless, we refer to this as a reference in the following discussion. 

In what follows, we find that when the mass inflow rate in the computational domain is substantially higher 
than the Eddington value ($\dot{M}_{\rm in}\gg \dot{M}_{\rm Edd}$), the simulation result with radiative diffusion 
is qualitatively similar to those without radiative diffusion because most of the photons are transported via advection 
with the flow rather than via diffusion. Moreover, since the radiative diffusion time becomes substantially shorter than 
the dynamical time, we need a long computational time to reach a steady state of the flow. Therefore, to avoid this 
issue, we first focus on simulations without radiative diffusion and discuss how the outer boundary conditions affect the 
flow properties (see \S \ref{subsubsec:fiducial} and \S \ref{subsubsec:All_BC}) and next show the case with radiative 
diffusion for comparison (see \S \ref{subsubsec:Diffusion}).

\subsection{Initial \& Boundary Conditions} \label{subsubsec:ICBC}

In this section, we present the initial and boundary conditions for our simulations. 
We set a computational region of $r_{\mathrm{min}}\leq r\leq r_{\mathrm{max}}$ and 
$0 +\epsilon \leq \theta \leq \pi - \epsilon$ with $r_{\mathrm{min}}= 3~r_{\mathrm{Sch}}$, 
$r_{\mathrm{max}}\simeq 1500~r_{\mathrm{Sch}}$ and $\epsilon=0.01$ to avoid a numerical singularity at the poles.
We assume a BH with mass of $M_{\rm BH}$ to be located at the coordinate center ($r=0$).
We set logarithmically spaced grids in the r-direction to capture the gas behavior at high density
and set uniformly spaced grids in the polar direction. 
The number of grid points is $\left(N_{r}, N_{\theta}\right)=(256,256)$, except for low resolution cases 
where $\left(N_{r}, N_{\theta}\right)=(128,128)$. 
We run six simulations with the same initial conditions but with different outer boundary conditions. 
In Table~\ref{tab:setups}, the simulation setups are summarized.

As initial conditions, we adopt power-law distributions of the gas density, pressure, and radial velocity 
that follow a Bondi-like inflow; namely,
$\rho(r)=\rho_{\rm out}(r/r_{\rm max})^{-3/2}$,
$p(r)=p_{\rm out}(\rho/\rho_{\rm out})^{4/3}$, where the adiabatic index is assumed to be $\gamma=4/3$, 
$v_r(r)=v_{\rm out}(r/r_{\rm max})^{-1/2}$,
and $v_\theta = 0$.
The normalization factors of those profiles are set to $\rho_{\rm out}=7.5\times10^{-9}$ g~cm$^{-3}$, 
$p_{\rm out} = 2\times 10^8~{\rm erg~cm}^{-3}$, $v_{\rm out} = -5\times 10^8~{\rm cm~s}^{-1}$.
This choice is adopted so that the mass inflow rate from $r_{\rm max}$ becomes as high as 
$\sim O(100-1000)~\dot{M}_{\rm Edd}$ and the flow is highly optically thick.
We add gas rotation to this Bondi-like profile, assuming the initial specific angular momentum 
to be the Keplerian value of $j_{z}=\sqrt{GM_{\rm BH}r}$, at $R(\equiv r\sin \theta) \geq \epsilon r_{\rm max}$; otherwise no initial rotation.
Note that the initial distribution of the rotational velocity does not affect the flow properties in the late time owing to 
angular momentum injection through the outer boundary (see below).

We impose injection boundary conditions at $r=r_{\rm max}$, unlike previous studies where the BH is fed 
by a dense torus that is initially set at the vicinity of the BH \citep[e.g.][]{jiang2014global,skadowski2015global}.
In our simulations, we consider four different outer-boundary conditions as summarized in Table~\ref{tab:setups}.
First, we divide the outer boundary layer into an equatorial ($|\theta - \pi/2|< \theta_{0})$ and 
a polar region ($|\theta - \pi/2|> \theta_{0}$), where $\theta_0$ is a parameter to be determined in each case.
In the polar regions, zero gradients across the boundary are imposed on physical quantities of the inflowing gas using 
ghost cells (numerically, boundary conditions are set two cells outside the grid referred to as ghost cells) and inflows are prohibited (i.e., $v_r \geq 0$ is imposed).
In the equatorial region, we set two different boundary conditions depending on the direction of the radial 
velocity. Specifically, the zero-gradient boundary conditions are imposed as in the polar region if $v_r \geq 0$ (i.e., outflows), and 
constant values of the density ($\rho_0$) and rotational velocity ($v_{\rm rot}$) are set at the outer boundary if $v_r < 0$ (i.e., inflows).
This treatment allows us to maintain a high accretion rate through the outer boundary when the gas is inflowing.
For Case-L (Low), we adopt $\theta_0 = \pi/4$, $\rho_{0}=6\times 10^{-9}~{\rm g~cm}^{-3}$, and $v_{\rm rot}=5\times 10^8~{\rm cm~s}^{-1}$.
For Case-N (Narrow), we set a smaller injection angle of $\theta_0 = \pi/12$. 
For Case-H (High), we increase the inflow density from the outer boundary by a factor of ten.

As for ``Injection" boundary conditions, the polar region conditions are the same as for the ``Low" boundary condition, while in the equatorial 
region, the gas is constantly injected into the domain with fixed density ($\rho_{0}=6\times 10^{-9}~{\rm g~cm}^{-3}$ ), radial velocity ($v_r=-3\times 10^8~{\rm cm~s}^{-1}$) and 
rotation velocity ($v_{\rm rot}=5\times 10^8~{\rm cm~s}^{-1}$). 
As for inner boundary conditions, sink boundary conditions are adopted since the inner radius is roughly the ``inner-most stable circular orbit" (ISCO)
radius. For the boundaries near the polar axes, periodic boundary conditions are applied to the rotational 
velocity and reflection boundary conditions are imposed to all the other physical quantities. Since we do not include cooling/heating or chemical 
reactions, the Courant time (the Courant number is set to 0.3) is sufficiently accurate to solve the hydrodynamical equations.

\section{2D Hydrodynamical simulations}\label{sec:results}

In this section, we present the simulation results with different boundary conditions. 
For all cases, we set $M_{\mathrm{BH}}=10^3~M_{\odot}$. 
In \S\ref{subsubsec:fiducial}, we show the results for Case-H (regarded as the``Fiducial" case) and discuss the general behaviour of 
gas dynamics. 
In \S\ref{subsubsec:All_BC}, we discuss how the different outer boundary conditions which provide different gas 
supply rates affect the properties of the accreting gas and the BH feeding rate,
In \S\ref{subsubsec:Diffusion}, we show the results with radiative diffusion in the dense, optically thick inflows
and discuss the effect on supercritical accretion. We also discuss the luminosity as well as photon-trapping effects.

\begin{figure}[t!]
\centering
\includegraphics[scale=0.7]{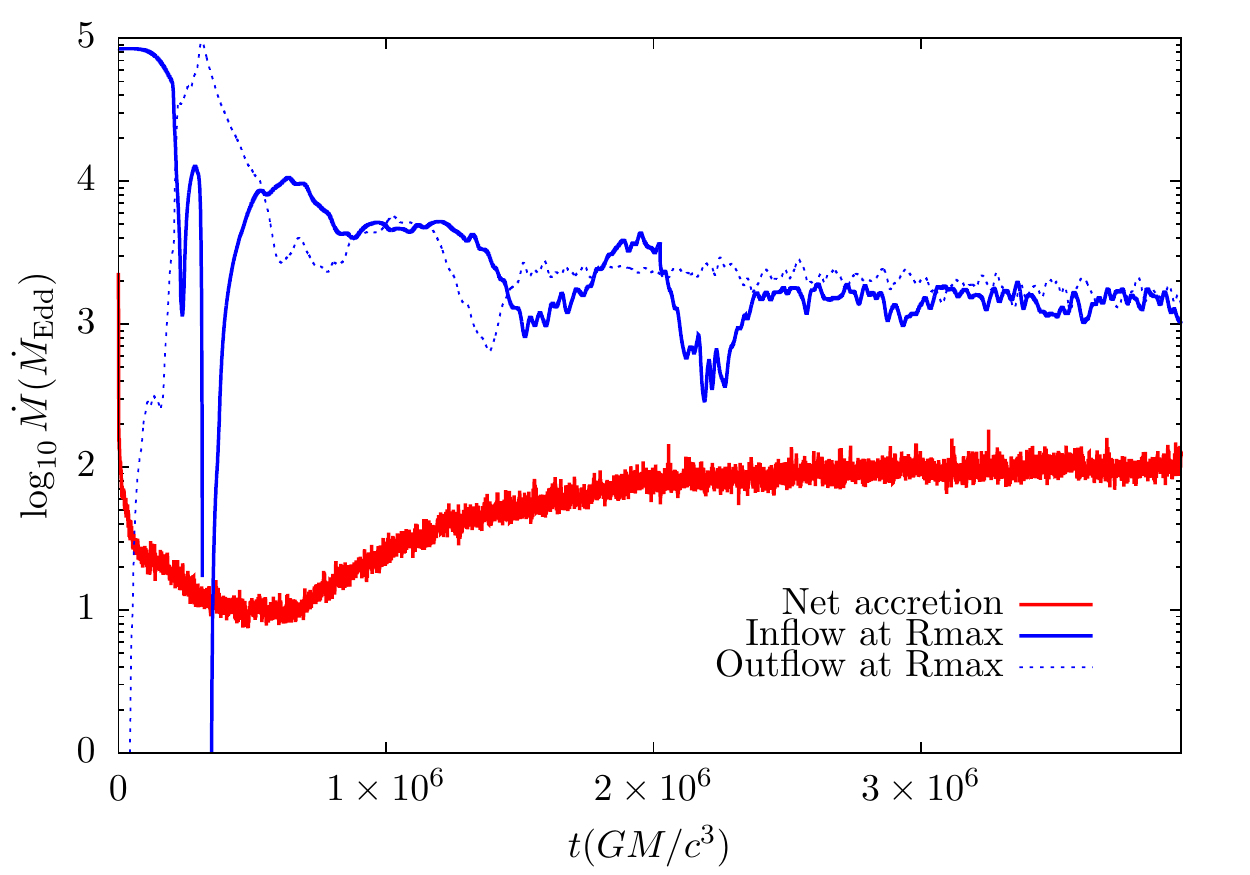}
\caption{Time evolution of the accretion rate onto the BH (red curve), the mass inflow (blue solid curve), and the outflow rate (blue dotted curve)
from the outer boundary for the fiducial case (the Case-H).
The accretion rates and timescales are normalized by the Eddington accretion rate and $t_0$, respectively. 
The simulation settles down to a quasi-steady state at $\dot{M}_{\rm BH}\simeq 100~\dot{M}_{\rm Edd}$ 
after $t \ga 2.3\times 10^6~t_0\simeq 1.2\times10^4~{\rm s}~(M_{\rm BH}/10^3~M_\odot)$.}
\label{fig:acc_evo_fid}
\end{figure}

\subsection{The Fiducial Case} \label{subsubsec:fiducial}

Fig.~\ref{fig:acc_evo_fid} shows the time evolution of the mass flux crossing the inner and outer boundaries
in the fiducial case. 
The curves indicate the accretion onto the BH (red solid), and the inflow and outflow rates at the outer boundary
(blue solid and dotted, respectively).
While in the early stage, both the inflow and outflow rate at the outer boundary oscillate,
they approach a constant value of $\sim2000~\dot{M}_{\mathrm{Edd}}$ at $t\ga 2.3\times 10^6~t_0$.
On the other hand, the accretion rate onto the BH becomes as small as $\sim 100~\dot{M}_{\rm Edd}$ in the quasi-steady state.

\begin{figure}[t!]
\centering
\includegraphics[scale=0.7]{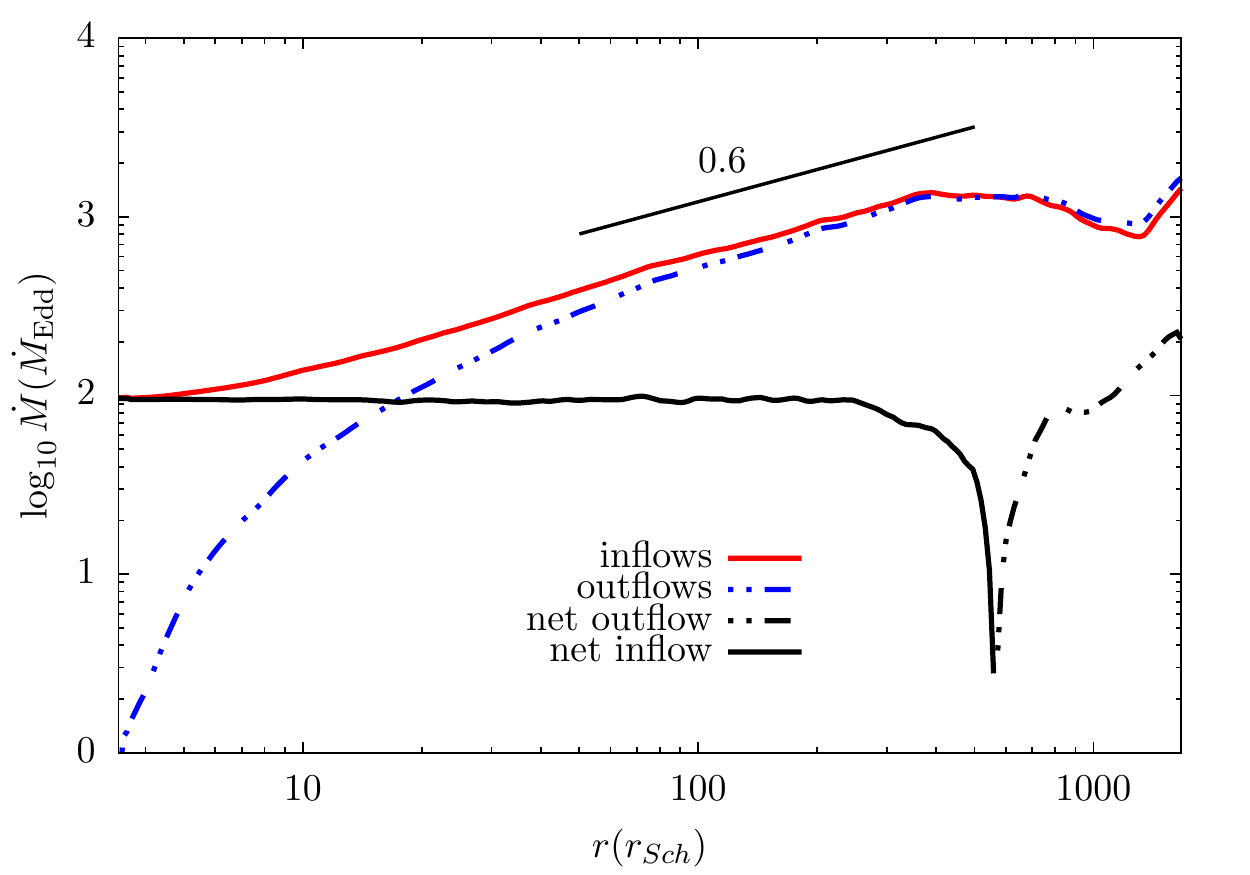}
\caption{Time-averaged radial profiles of the net accretion rate (black), the mass inflow rate (red), and the mass outflow rate (blue)
in the fiducial case (the Case-H).
These profiles are time-averaged over $3\ \la t/(10^6~t_0) \la 4$. 
While the inflow rate balances the outflow rate at larger radii, the inflow rate dominates
at the vicinity of the BH.
As a result of mass removal by outflows, the inflow rate decrease toward the center following $\dot{M}_{\rm in}(r)\propto r^{0.6}$,
and $5\%$ of the inflowing gas from the outer boundary feeds the BH.}
\label{fig:mass_flow_fid}
\end{figure}

Fig.~\ref{fig:mass_flow_fid} shows the time-averaged radial structure of the angle-integrated mass inflow (red) and outflow (blue) rates in the fiducial case 
when the system is in a quasi-steady state ($3\la t/(10^6~t_0)\la4$).
These rates are defined by
\begin{equation}
\dot{M}_{\mathrm{in}}=-2 \pi r^{2} \int_{0}^{\pi} \rho \min \left(v_r, 0\right) \sin \theta \mathrm{d} \theta,
\end{equation}
and 
\begin{equation}
\dot{M}_{\mathrm{out}}=2 \pi r^{2} \int_{0}^{\pi} \rho \max \left(v_r, 0\right) \sin \theta \mathrm{d} \theta,
\label{eq:mout}
\end{equation}
where the net accretion rate is calculated by $\dot{M}_{\rm net}=\dot{M}_{\rm in}-\dot{M}_{\rm out}$ with mass loading factor 
$\beta=|\dot{M}_{\rm out}(r=r_{\rm max})/\dot{M}_{\rm BH}|$ and $\dot{M}_{\rm BH}=\dot{M}_{\rm net}(r=r_{\rm min})$.
Note that $\dot{M}_{\rm out}$ in Eq.~(\ref{eq:mout}) is not necessarily the same as the outflow rate of gas escaping to infinity, since it includes all gas with $v_r>0$.
At larger radii ($r>30~r_{\rm Sch}$), the inflow and outflow rates are almost balanced and thus 
the inflow rate decreases toward the center following $\dot{M}_{\rm in}(r) \propto r^{0.6}$.
Within $r \simeq 20-30~r_{\rm Sch}$, the outflow ceases and the inflowing gas dominates.
As a result, the accretion rate is reduced significantly from the inflow rate at $r\simeq r_{\rm max}$ to $100~\dot{M}_{\rm Edd}$.
We note that the net accretion profile is constant within $\sim 300~r_{\rm Sch}$\footnote{Outside this radius, a fountain-like 
structure forms in the accretion flow and thus the flow does not settle down completely yet. 
The time-dependent fountain-like motion carries a small fraction ($\sim 3\%$) of the total gas mass 
outward from the domain in a $10^6 ~t_0$ duration. 
Nevertheless, the flow properties in the quasi-steady state would not be much different from those we discuss here.}, 
indicating that the gas over a wide range of the spatial scales sufficiently approaches a quasi-steady state.
The reduction of the inflow rate (i.e., $\dot{M}_{\rm in} \propto r^p; p>0$) is generally seen in most numerical simulations 
of radiatively inefficient accretion flows (RIAF, $p\sim 0.75-1$; \citealt{stone1999hydrodynamical,Igumenshchev1999}).
The power-law index $p\simeq 0.6$ is smaller than that predicted in the convection-dominated accretion flow (CDAF) model
\citep[][]{Narayan2000,Quataert2000,Abramowicz2002,Igumenshchev2002,Yuan2014}. The value of $p$ can be used to characterize 
the outflow strength. In different accretion models, the outflow strength has been found in the range $0.5\la p\la 1$.

\begin{figure*}
\begin{center}
\includegraphics[scale=0.8]{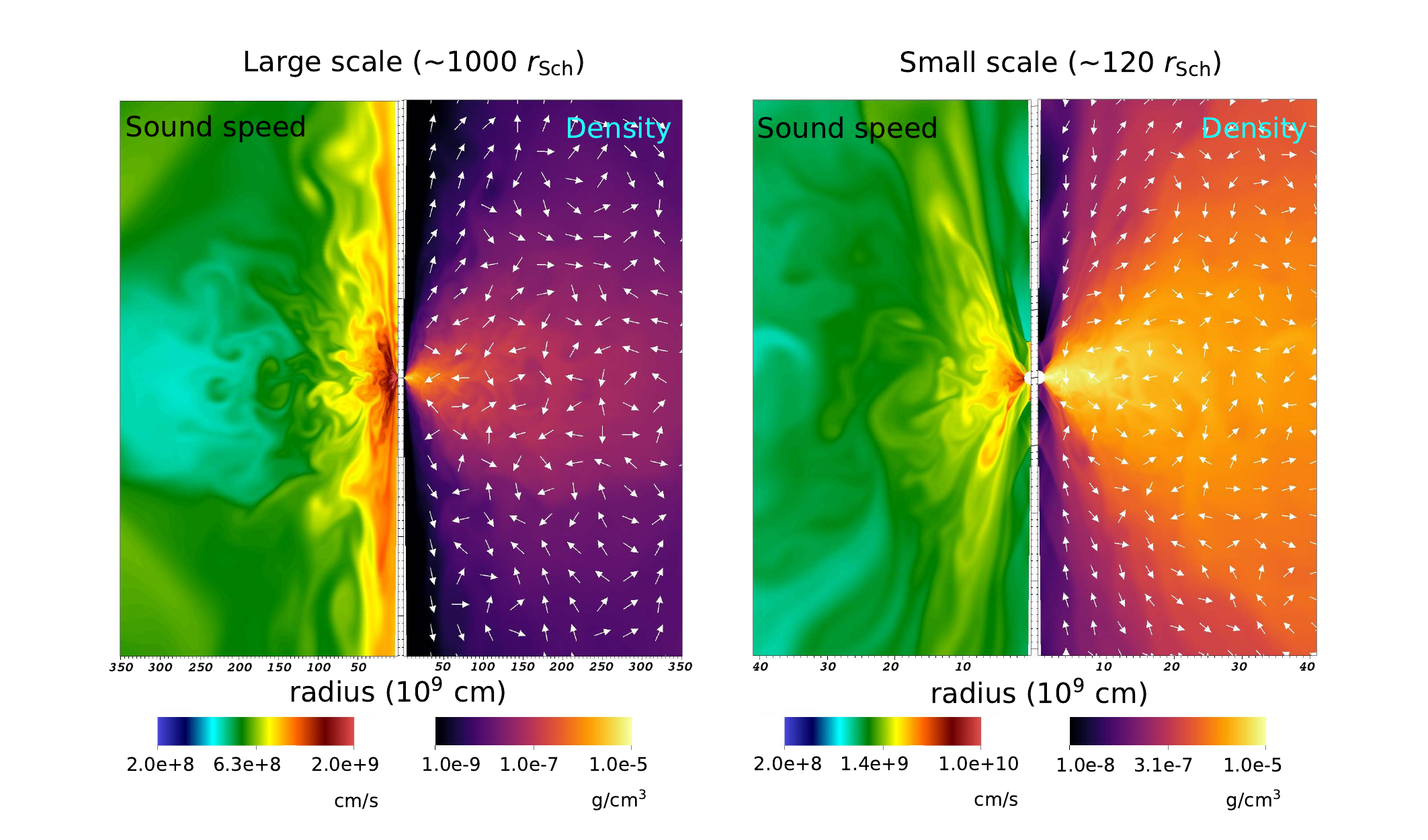}
\caption{ 2D sound speed ($\equiv \sqrt{\gamma p/\rho}$) distribution (left part), and 2D density distribution (right part) in Case-H with 
velocity vectors shown when the system is in a quasi-steady state ($t=3.8\times 10^6~t_0$).
{\it Left panel}: Large scale distribution. {\it Right panel}: small-scale distribution. 
The gas is outflowing in the polar region on large scales(see velocity vectors in the left panel), while near the equator, the gas is highly turbulent. 
The radiation is trapped near the BH, with a small fraction leaking in the polar direction, indicated by the high sound speed along the 
polar axis.  }
\label{fig:2D-plots}
\end{center}
\end{figure*}

\begin{figure*}
\includegraphics[scale=0.24]{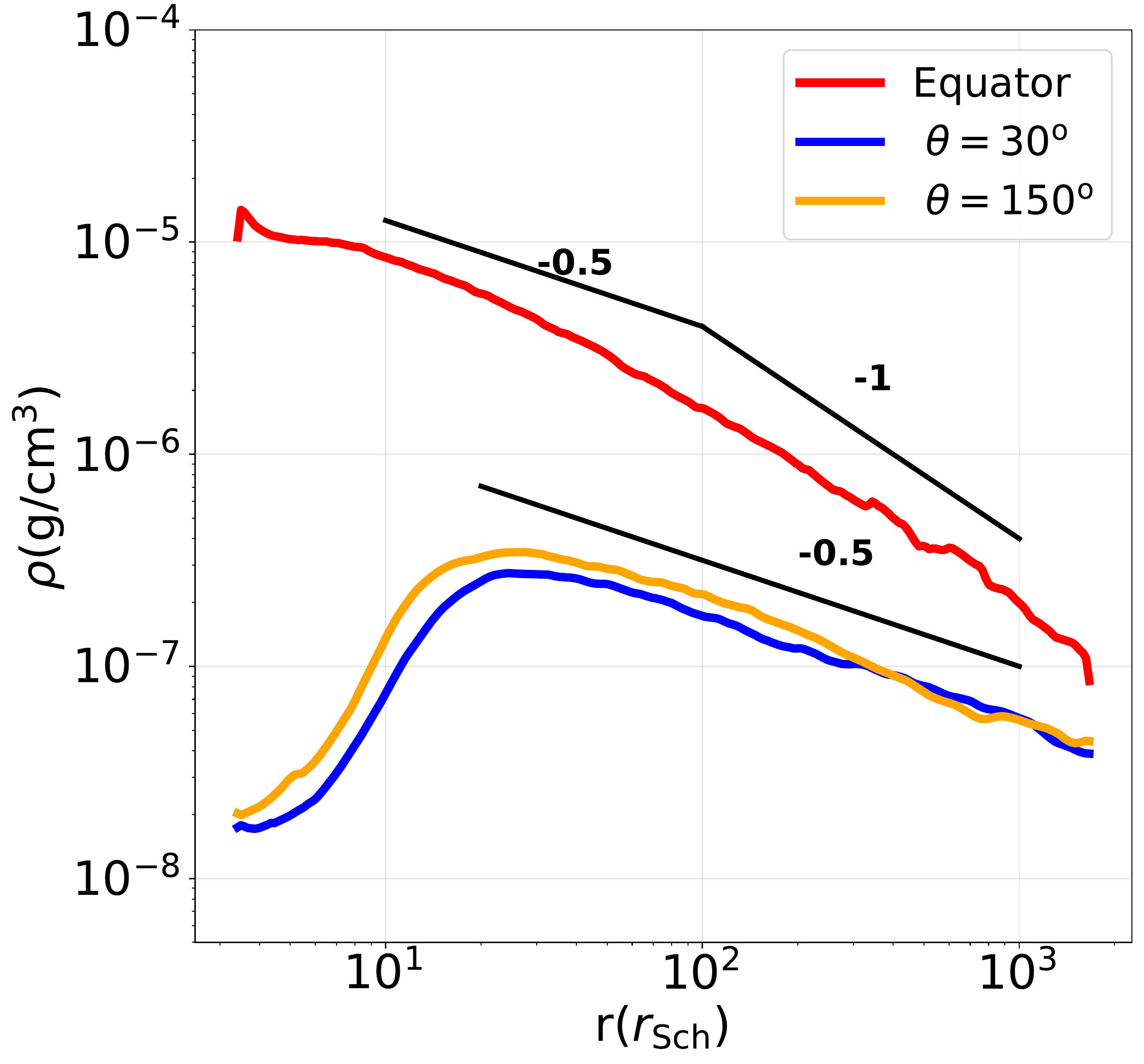}\hspace{2mm}
\includegraphics[scale=0.24]{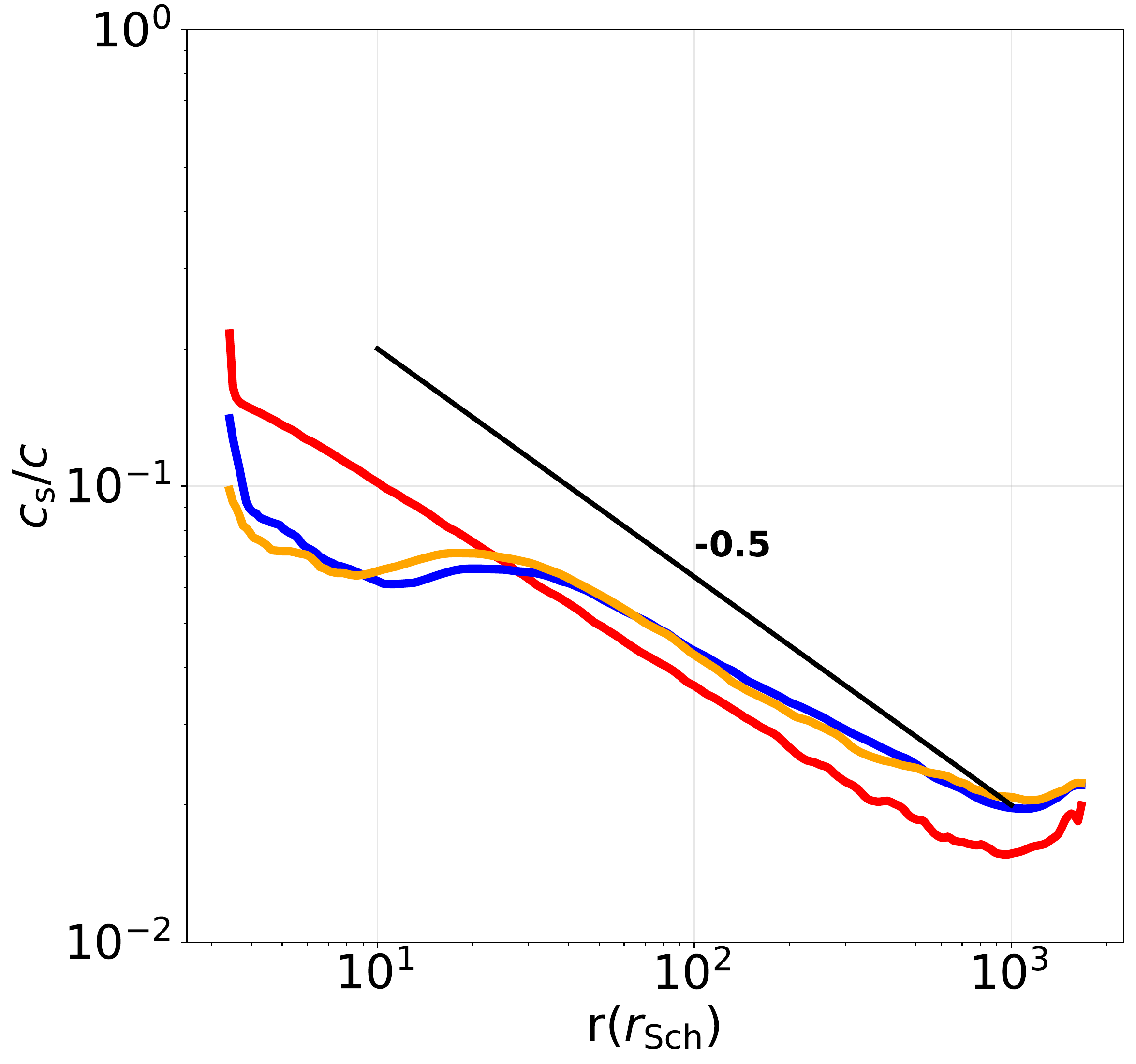}\hspace{2mm}
\includegraphics[scale=0.24]{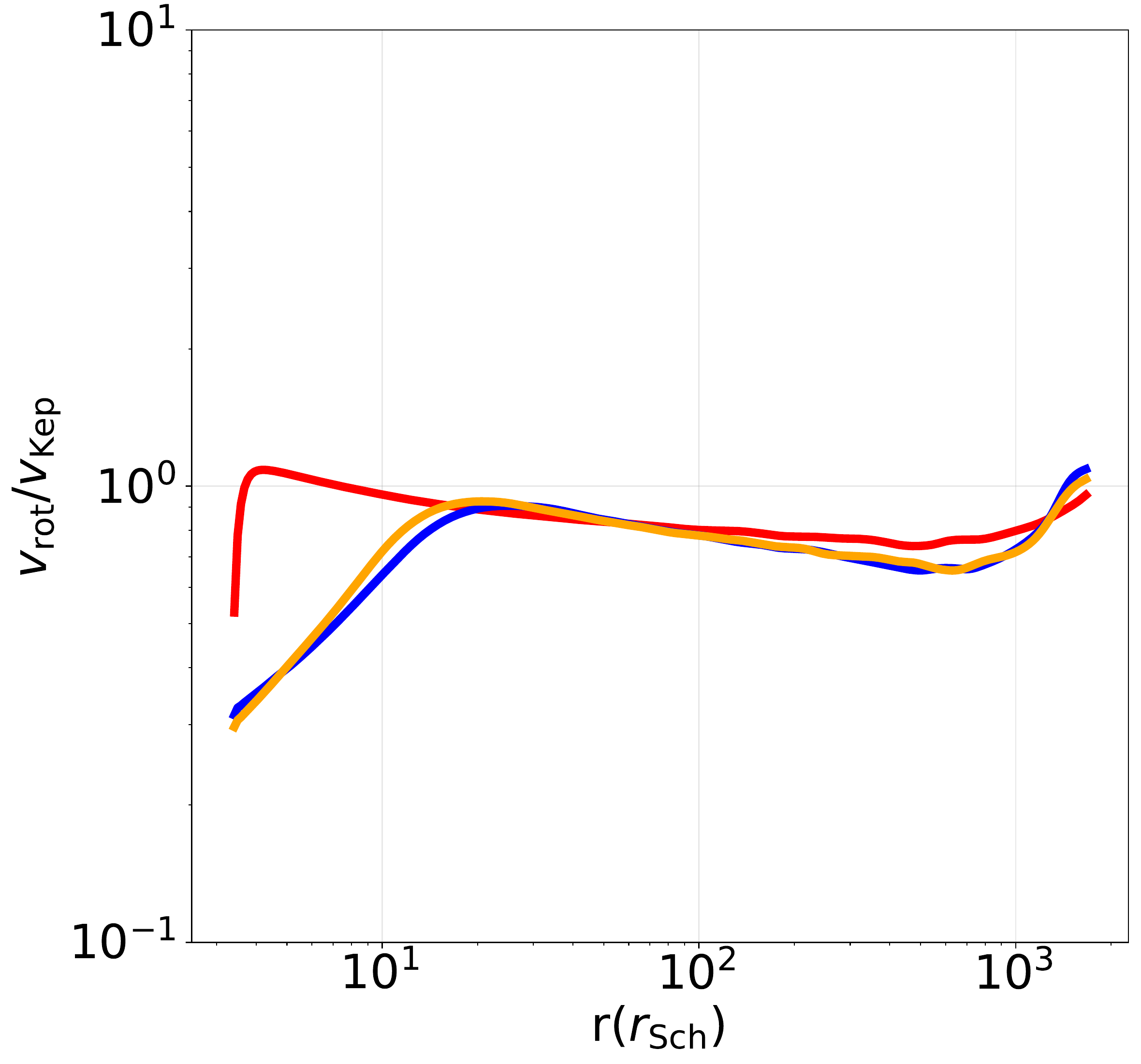}
\caption{Time-averaged profiles of the density (left), sound speed (middle), and rotation velocity (right)
along the equator (red) and the directions at $\theta=30^\circ$ (blue) and $150^\circ$ (orange), respectively. 
The density profile along the equator is characterized by a broken power-law profile that follows $\rho \propto r^{-1/2}$ in the inner region
of $r<30~r_{\rm Sch}$ and $\rho \propto r^{-1}$ in the outer region.
The former slope agrees with that expected in a CDAF solution \citep{Quataert2000}.
The accretion flow on the equatorial plane is rotationally (pressure) supported inside 
(outside) $r\sim100~r_{\rm sch}$, but the gravitational force exerted by the central BH marginally 
dominates at all radii over the sum of the pressure-gradient force and centrifugal force.}
\vspace{10pt}
\label{fig:profiles_fid}
\end{figure*}

Fig.~\ref{fig:2D-plots} shows the two-dimensional distribution of the sound speed and gas density at two different 
scales for the fiducial case when the simulation terminates. 
On the larger scale (left panel), the overall flow pattern is governed by convective motion and shows that the gas 
is inflowing through the equatorial region and a similar amount of gas is outflowing to the polar regions. On the 
smaller scale (right panel), the inflows dominate over the outflows owing to strong gravity.
The sound speed at the inner and polar region is higher than 
elsewhere indicating that radiation is trapped in the inner region but can escape through polar region and heat the ambient gas. 
The accretion flow becomes highly turbulent due to energy generated via viscosity. This behavior is consistent with
those found in simulations of RIAFs \citep{stone1999hydrodynamical,Igumenshchev1999}. 
However, in reality, turbulent motion is essentially a three-dimensional (3D) effect that would break the axisymmetric flow 
structure imposed in our simulations \citep[but see also][]{Igumenshchev2000}. 
Exploring the nature of multi-dimensional effects is left for future investigation.

Fig.~\ref{fig:profiles_fid} presents the time-averaged radial profiles of the density (left), sound speed (middle), and rotational velocity (right). 
The sound speed and rotational velocity is normalized by the speed of light and the Keplerian velocity, respectively.
The three different curves show the profiles along three different directions: $\theta =30^\circ$ (blue), $\theta =90^\circ$ (equator; red), and 
$\theta =150^\circ$ (orange).
The density profile along the equator follows a power-law of $\rho \propto r^{-1/2}$ 
at $r\la 30~r_{\rm Sch}$ and agrees with those expected in a CDAF solution \citep{Quataert2000}, while 
the slope becomes as steep as $\rho \propto r^{-1}$ in the outer region.
Strong and fast outflows driven by the radiation pressure gradient force
create low-density cavities in the bipolar directions, where the 
outflowing matter obeys $\rho \propto r^{-1/2}$ (for $>30~r_{\rm Sch}$).
The sound speed profile is approximated by a single power-law of $c_{\rm s}\propto r^{-1/2}$ for all the directions,
indicating that the pressure gradient force partially balances with the BH gravity; namely the ratio is $<0.3$ at $r\lesssim 100~r_{\rm Sch}$.
The rotational velocity is close to the Keplerian value and the ratio of $v_{\mathrm{rot}}/v_{\mathrm{Kep}}\sim 0.87$ is consistent with 
the analytical expression for $\gamma=4/3$ \citep{Quataert2000}
with a gradual increase toward the center. Indeed, the accretion flow on the equatorial plane is rotationally (pressure) supported 
inside (outside) $r\sim100~r_{\rm Sch}$, but the gravitational force exerted by the central BH marginally dominates at all radii over 
the sum of the pressure-gradient force and centrifugal force.

\begin{figure*}
\begin{center}
\includegraphics[scale=0.3]{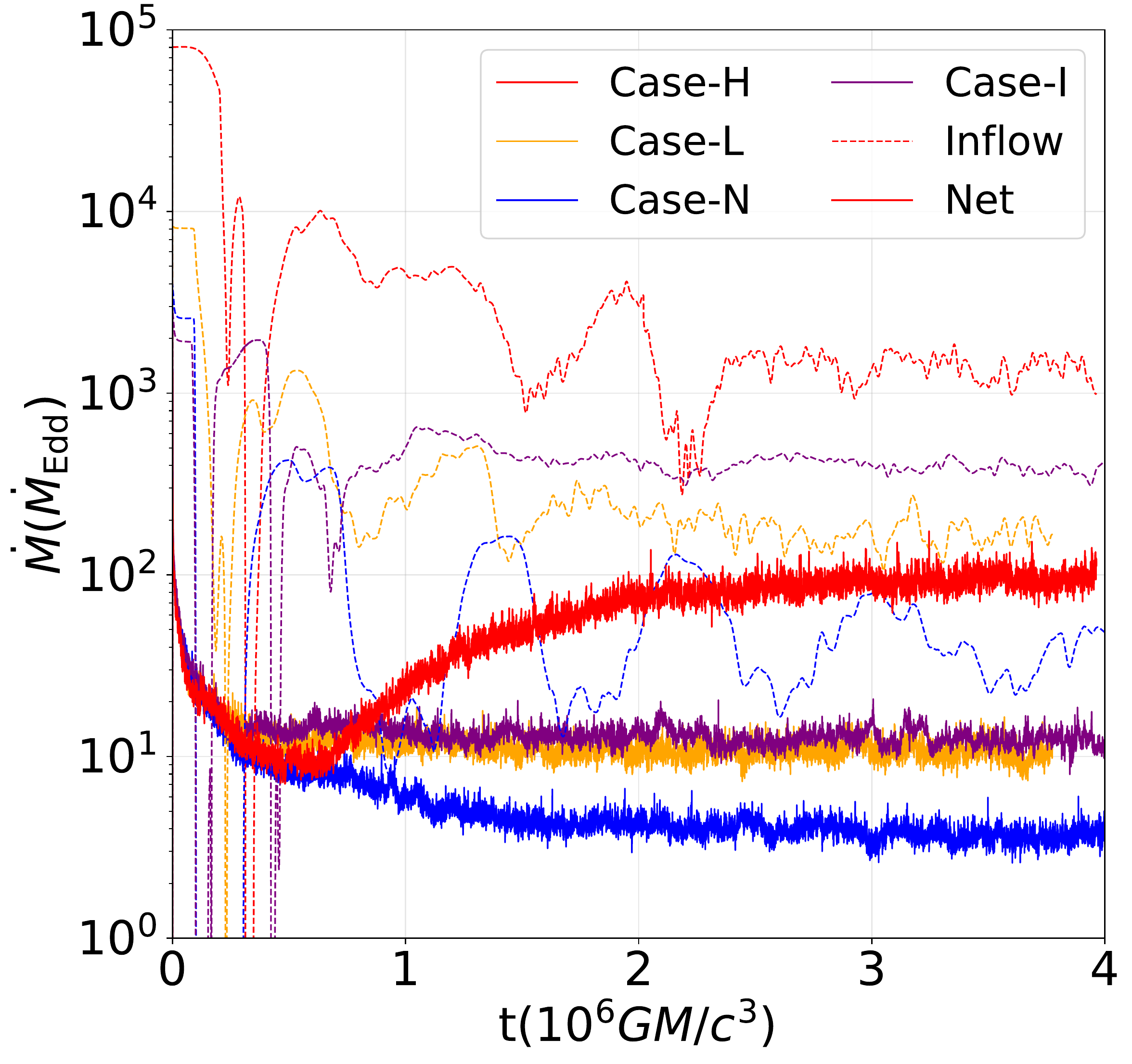}\hspace{10mm}
\includegraphics[scale=0.3]{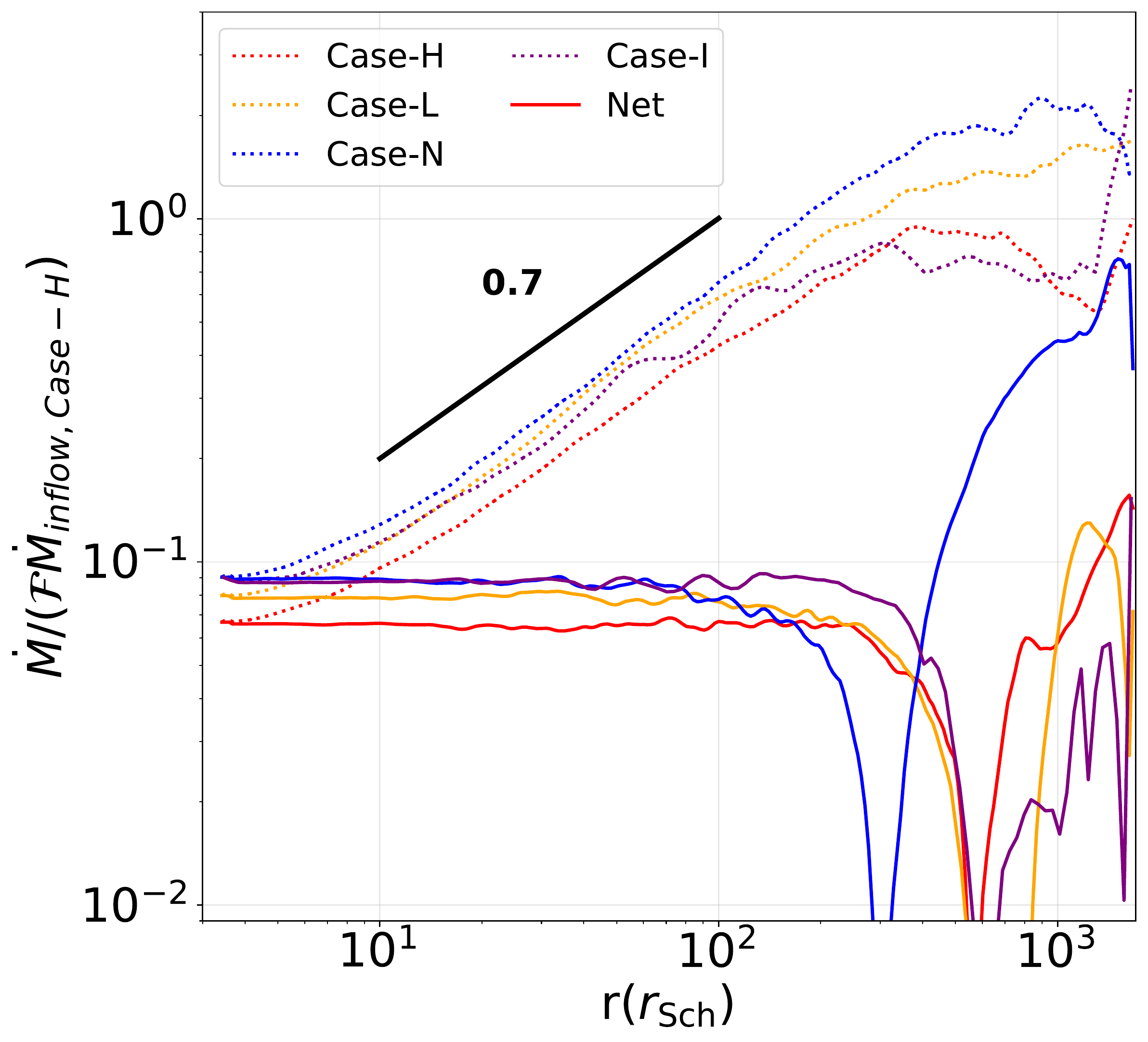}
\caption{{\it Left penal} : Time evolution of the BH accretion rate (solid) and the mass inflow rate at the outer boundary (dotted) 
in the four cases with different outer boundary conditions: Case-H (red), Case-L (orange), Case-N (blue), Case-I (purple). 
In each case, the system reaches a quasi-steady state after $t\sim 2.3\times10^6~t_0$.  
{\it Right panel} : Time-averaged radial profiles of the net BH accretion rate (solid) and the mass inflow rate (dotted) for the four cases.
These profiles are normalized by the inflow rate at the outer boundary in the fiducial Case-H and a scale factor $\mathcal{F}\equiv \rho_{{\rm 0}}
v_{r}\Omega/(\rho_{\rm 0, Case-H}v_{r,{\rm Case-H}}\Omega_{\rm Case-H})$, which is the ratio of the density ($\rho_0$), 
velocity ($v_r$) of the gas injected at the outer boundary, and solid angle ($ \Omega$) of the injection region in each case relative to the fiducial case.
Namely, $\mathcal{F}=10.0$ (Case-L), $27.7$ (Case-N), and $9.2$ (Case-I), respectively.
The normalized profiles exhibit the self-similar behavior of radiation-dominated accretion flows with 
$\dot{M}_{\rm in}\propto r^{0.7}$.
}
\label{fig:acc_all}
\end{center}
\end{figure*}

In summary, due to the presence of strong outflows driven by the pressure gradient force, the inflow rate decreases toward the center as $\dot{M}_{\mathrm{in}}\propto r^{0.6}$
and thus the net accretion rate onto the BH is significantly reduced from the gas supply rate injected at the outer boundary. 
Since the efficiency of mass accretion defined by $\eta ~[\equiv \dot{M}_{\rm in}(r_{\rm min})/\dot{M}_{\rm in}(r_{\rm max})]$ is kept as high as 
$\eta \simeq 0.05$ (or mass loading factor $\beta\sim 19$, in our cases)
even under strong outflows, the BH feeding rate reaches $\simeq 100~\dot{M}_{\rm Edd}$ in the fiducial case.
We note that the global structure of the flow is found to settle down to this quasi-steady state after $t\ga 2.3\times 10^6~t_0$, which is $\sim 10-100$ times longer 
than the computational timescales addressed in previous (magneto-)radiation hydrodynamical simulation studies
\citep[e.g.][]{ohsuga2005supercritical,jiang2014global,skadowski2015global,jiang2019super}.

\begin{figure*}
\includegraphics[scale=0.24]{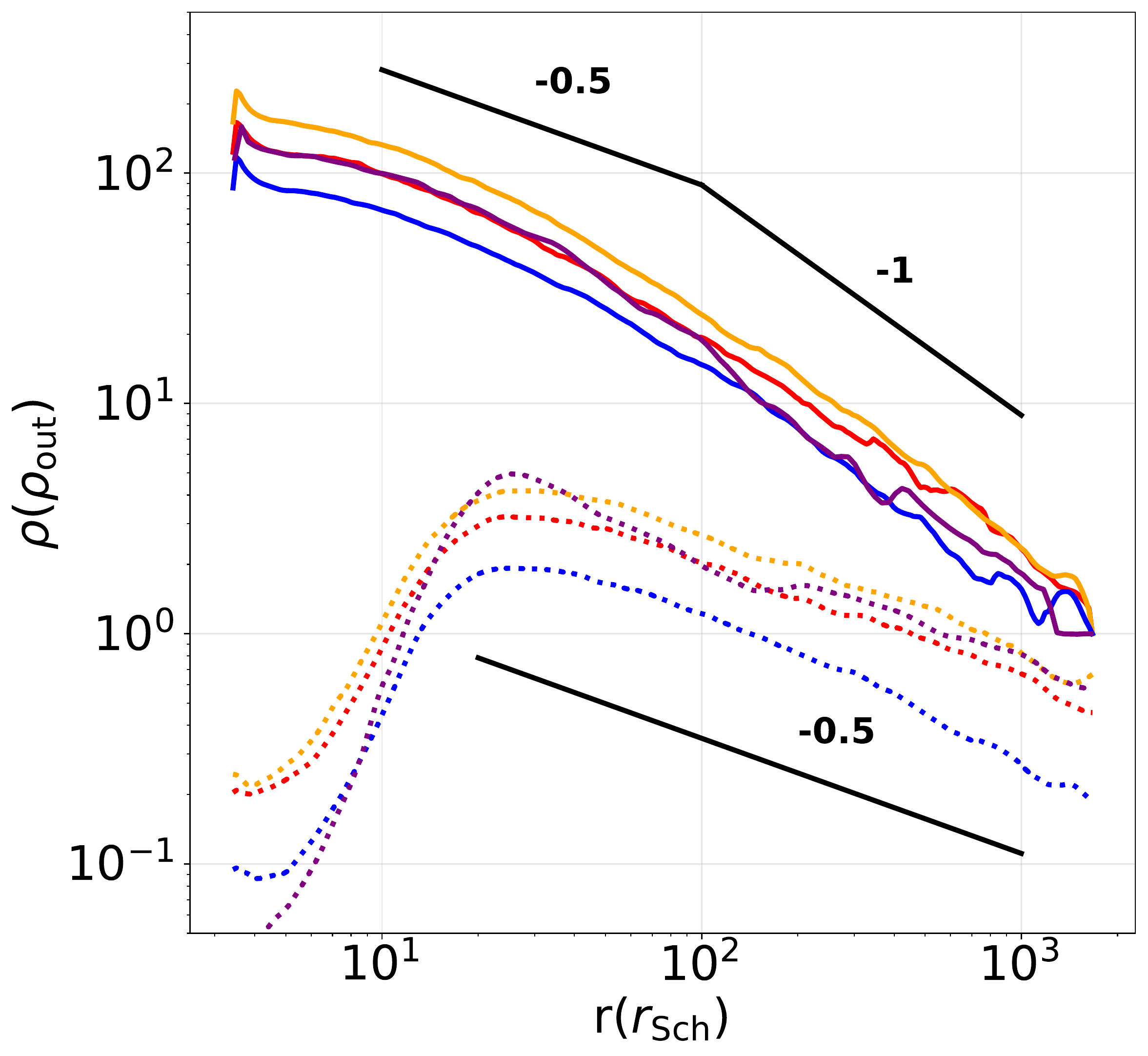}\hspace{2mm}
\includegraphics[scale=0.24]{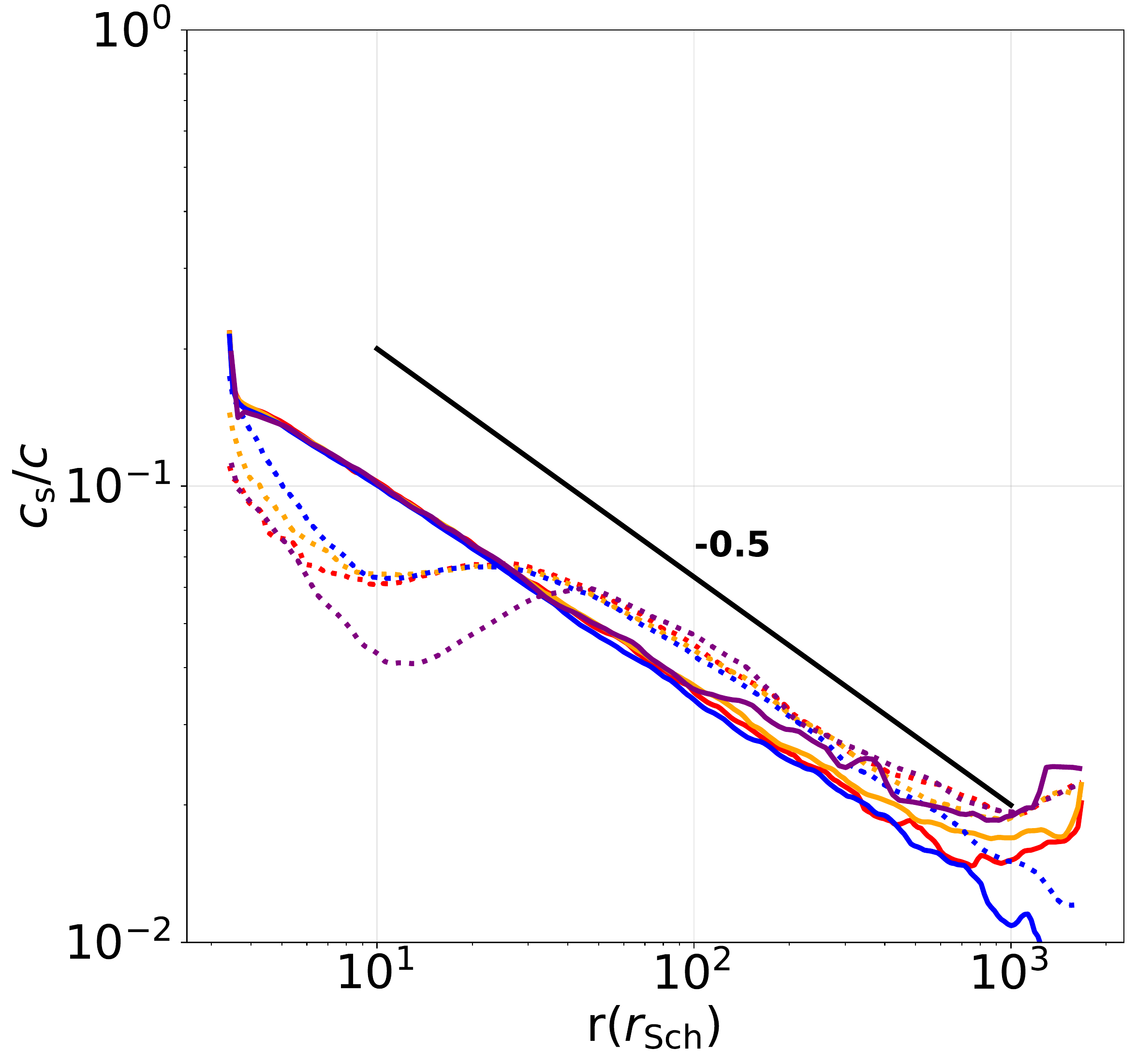}\hspace{2mm}
\includegraphics[scale=0.24]{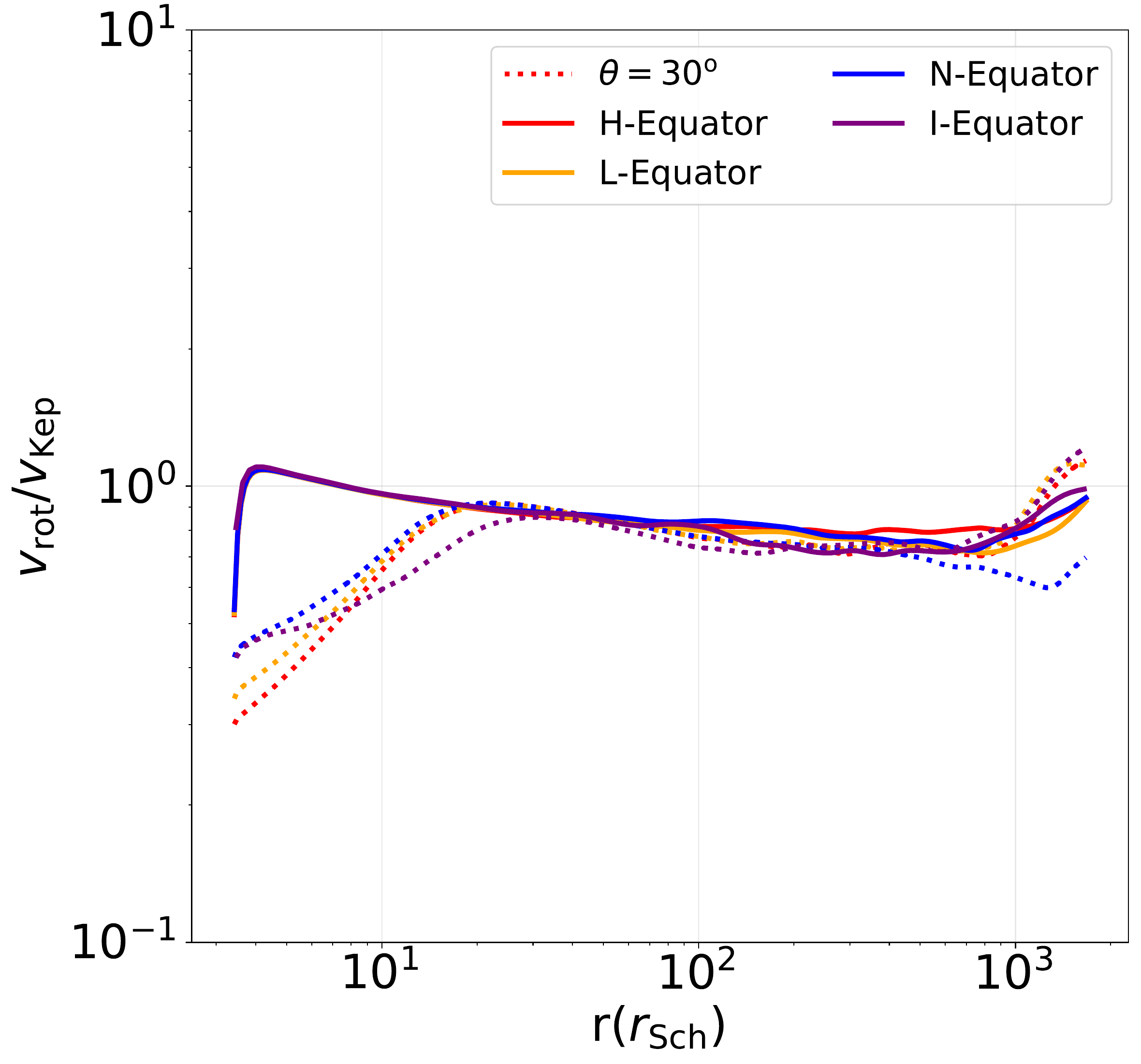}
\caption{Time-averaged radial profiles of physical quantities in the four cases: Case-H (red), Case-L (orange), Case-N (blue), and Case-I (purple). 
The solid curves represent the profiles along the equator, while the dotted curves are those along the polar angle $\theta =30^\circ$. 
Note that the density is normalized by that injected at the outer boundary.
In all cases, the profiles are similar to those of the fiducial case, without showing a new characteristic physical scale.}
\label{fig:profiles_all}
\vspace{5mm}
\end{figure*}

\vspace{2mm}
\subsection{Simulations with Different Boundary Conditions} \label{subsubsec:All_BC}

Next, we discuss the effect of the outer boundary conditions on the properties of the accretion flows.
In the left panel of Fig.~\ref{fig:acc_all}, we show the time evolution of the net accretion rate (solid) and the mass inflow rate (dotted)
for the four cases; the fiducial Case-H, Case-L, Case-N, and Case-I.
Depending on the outer boundary conditions, the BH accretion rates range over $\dot{M}_{\rm BH}\simeq 3-100~\dot{M}_{\rm Edd}$
and each of them approaches a constant value at $t \ga 2\times 10^6~t_0$.
The reduction factor of the net accretion rate compared to the fiducial case (Case-H) is consistent with the reduction in the inflow rate 
due to the differences at the outer boundary. 
Regardless of the different gas supply rates, the accretion efficiency for all cases is $\eta \sim 0.05$.
The overall behavior of the mass inflow rate from the outer boundary is also qualitatively similar 
except in Case-N, which shows periodic oscillations with larger amplitudes.
This is because a large-scale circulation crossing the equator disturbs the gas inflowing from $\sim r_{\rm max}$
through a narrow injection angle.

\begin{figure*}
\begin{center}
\includegraphics[scale=0.3]{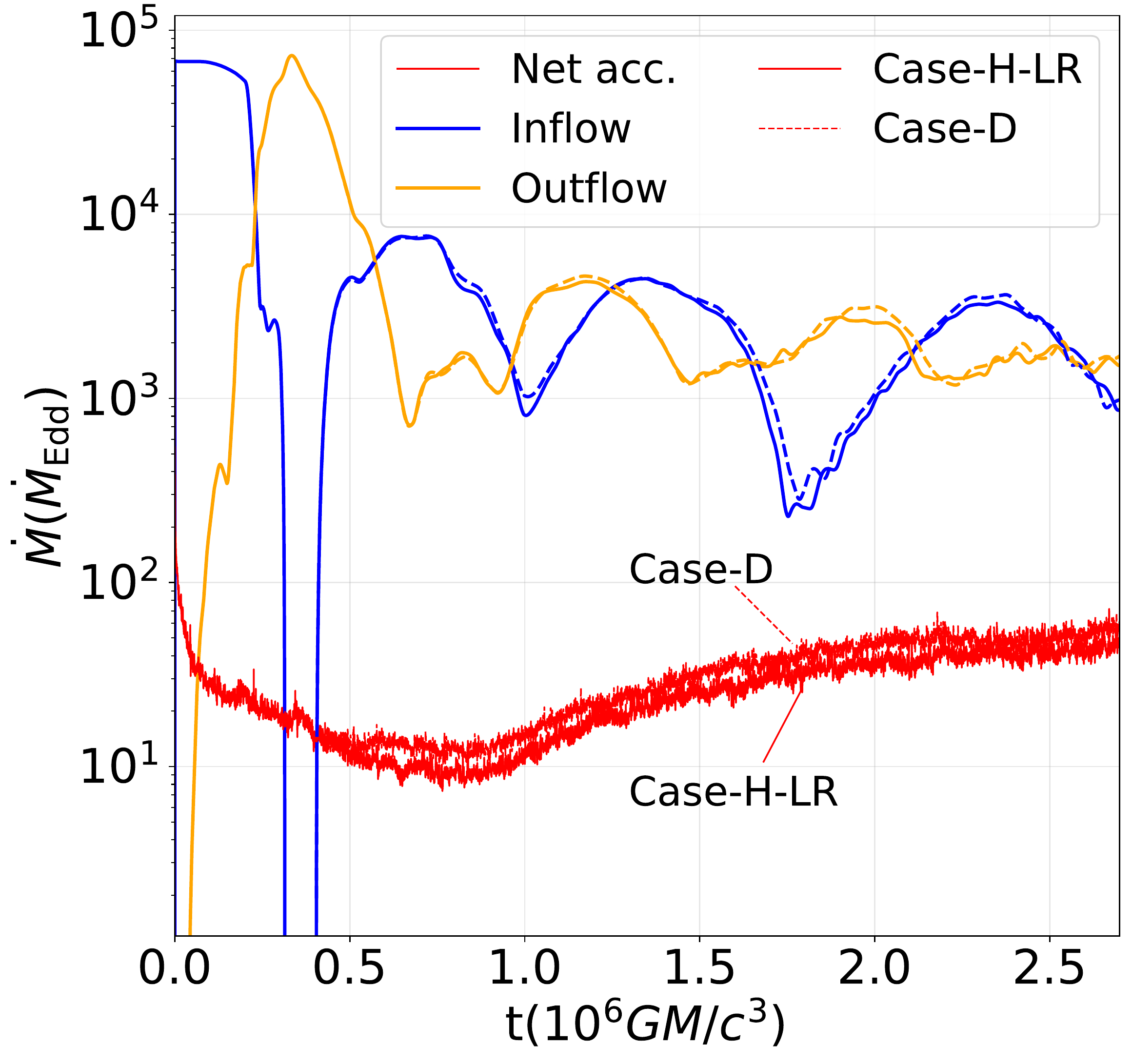}\hspace{10mm}
\includegraphics[scale=0.3]{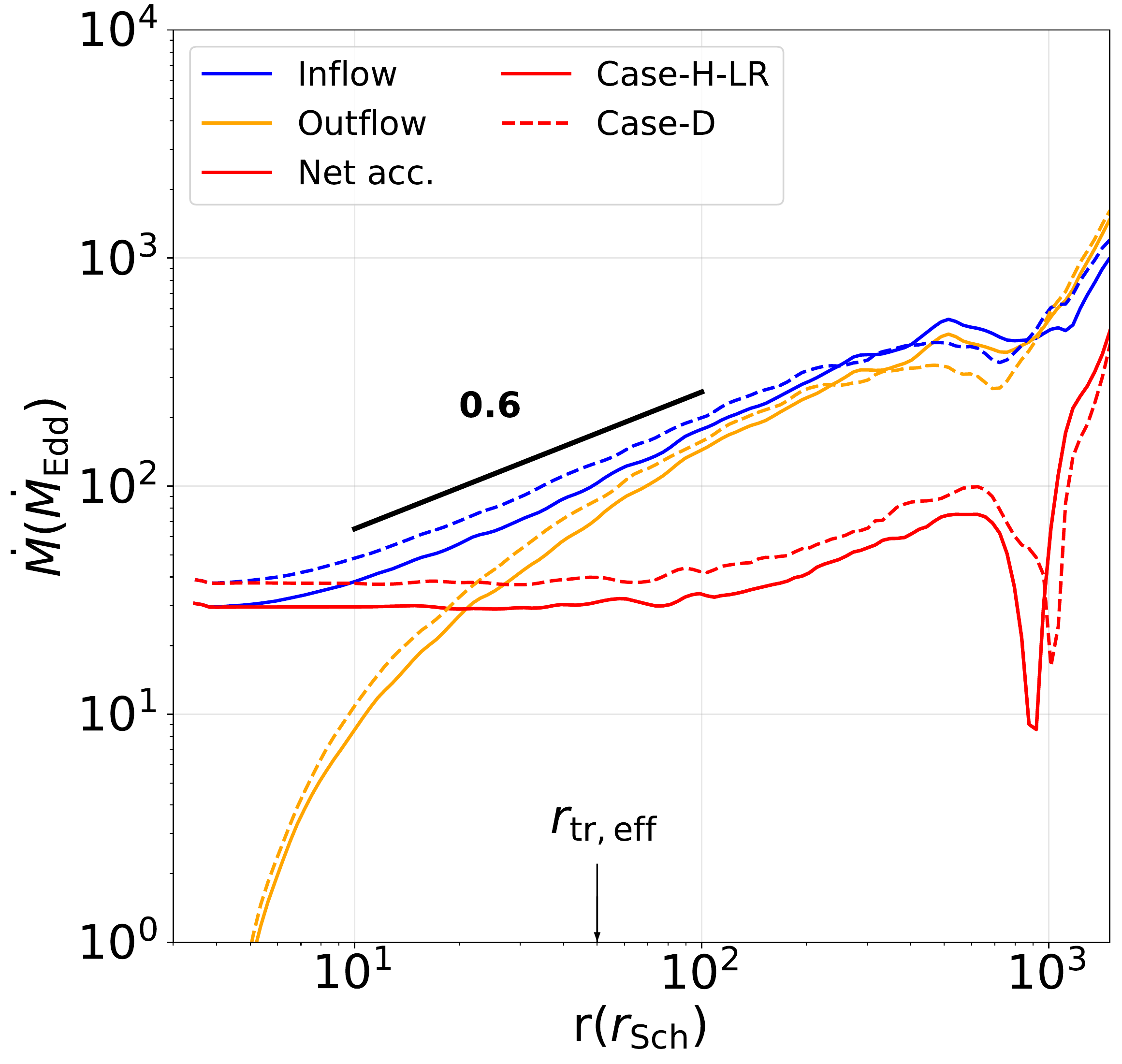}
\caption{Comparison between the cases with radiative diffusion (Case-D) and that without diffusion (Case-H-LR).
Note that the two cases adopt a lower resolution of ($N_r,N_\theta$) = (128,~128) and the latter case is identical to Case-H
except for the resolution. 
Overall, the mass flow rates in the two cases are similar, suggesting that energy transport 
via radiative diffusion plays a minor role in determining the bulk properties of highly optically thick accretion flows. 
The arrow in the right panel marks the angle-averaged trapping radius ($ r_{{\rm tr, eff}} \sim 50~r_{\rm Sch}$) inferred from rest-frame luminosity (see \S \ref{subsubsec:Diffusion}).
}
\label{fig:acc_diffusion}
\end{center}
\end{figure*}

In the right panel of Fig.~\ref{fig:acc_all}, we present the time-averaged radial profiles of the net accretion and mass inflow rates in all four cases.
These rates are normalized by the BH accretion rate in the fiducial case and a scale factor $\mathcal{F}\equiv \rho_{{\rm 0}}
v_{r}\Omega/(\rho_{\rm 0, Case-H}v_{r,{\rm Case-H}}\Omega_{\rm Case-H})$,
which is the ratio of the injection density, solid angle, and velocity in each case relative to the fiducial case.
Namely, $\mathcal{F}=10.0$ (Case-L), $27.7$ (Case-N), and $9.2$ (Case-I), respectively.
The radial profiles of these normalized rates clearly show the self-similar nature of the radiation-dominated accretion flows,
i.e., $\dot{M}_{\rm in}(r) \propto r^{p}$ with $p\sim 0.5-0.7$.
Similarly, Fig.~\ref{fig:profiles_all} shows the self-similarity of the gas density (left), sound speed (middle), and rotational velocity (right)
both along the equator (solid) and the polar angle $\theta=30^\circ$ (dotted).
Note that the density is normalized by that injected through the outer boundary.
Reflecting the scale-free nature of the flows, the profiles of these quantities are similar to those in the fiducial case, 
without introducing any new characteristic physical scale.

\begin{figure*}
\begin{center}
\includegraphics[scale=0.31]{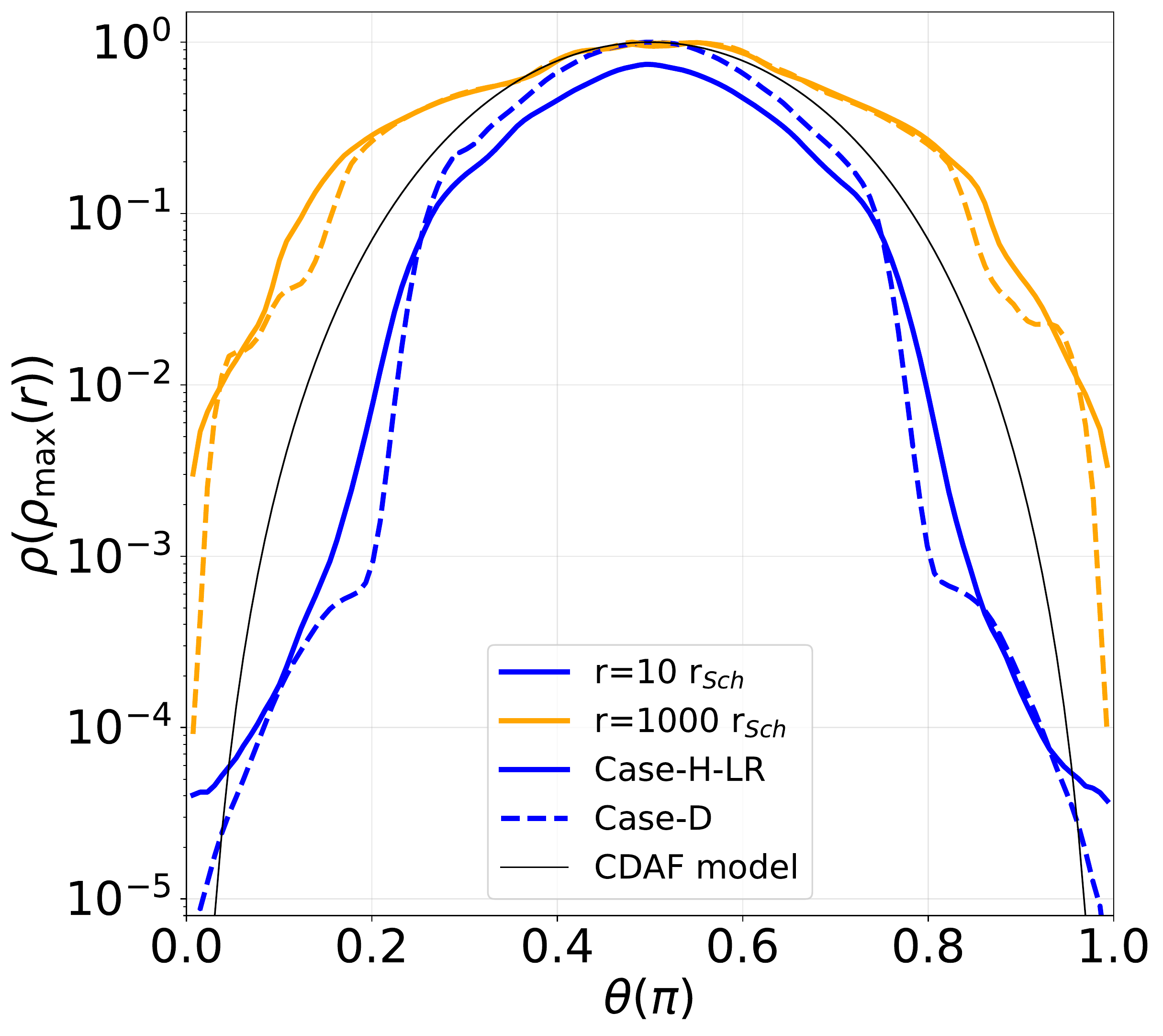}\hspace{8mm}
\includegraphics[scale=0.31]{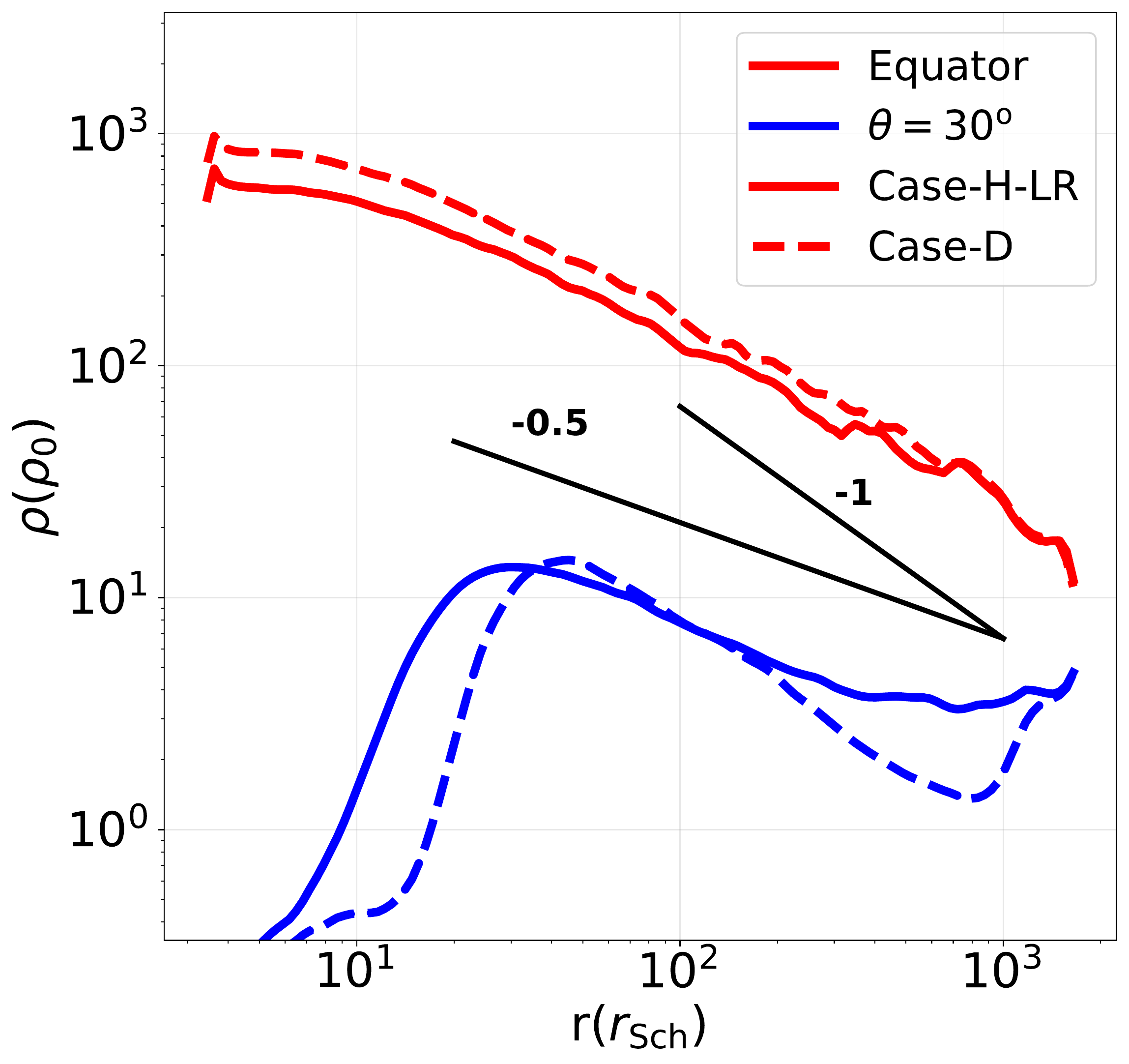}
\caption{
{\it Left panel} : Angular dependence of the density at two different radii for Case-D (solid) and Case-H-LR (dashed).
The profiles are normalized by the maximum density in each case.
The solid black curve presents that expected from a CDAF solution with $\gamma=4/3$ \citep{Quataert2000}.
{\it Right panel} : Radial profiles of the density along the equator (red) and the polar angle $\theta=30^\circ$ (blue)
for the two cases.
}
\label{fig:theta_profiles}
\end{center}
\end{figure*}

Recently, \citet{kitaki2021outflow} (hereafter, K21) conducted simulations similar to our Case-I run, 
where gas is injected at a constant inflow rate of $\sim 300~\dot{M}_{\rm Edd}$ through a narrow funnel 
around the equator ($\Delta \theta = 0.02~\pi$).
In their simulations, K21 found that almost all the gas accretes onto the BH with an accretion efficiency of $\eta \simeq 0.88$
and produces weak outflows with a mass loading factor $\beta \simeq 0.14$.
While the outflows are launched from the interior of the photon-trapping radius ($r_{\rm tr}\simeq 450~r_{\rm Sch}$ in their simulations), 
most of the outflowing gas fails to escape from the outer boundary and falls back onto the disk. 
In contrast, we find a large mass loading factor for the outflows with $\beta \simeq 20$ on large scales and thus
only a small fraction of the inflowing gas feeds the BH.  
The large discrepancy in the bulk properties of the flow and mass loading factor is attributed to the following.
First, K21 adopted a computational domain that cover a hemisphere and impose equatorial mirror-symmetry of 
the accretion flow across the equatorial plane.
The imposed symmetry tends to suppress global convective motion that crosses the equator and is known to produce 
outflows coherently through the equator rather than the polar regions \citep[see][]{Li2013}.
Under K21's setup, the equatorial outflow dissipates its kinetic energy by forming shocks at the interface with injected gas,
which pushes the gas inward, and thus the production of outflows is suppressed. Secondly, they assume a larger viscous 
parameter of $\alpha=0.1$, which is known to cease convective/turbulent motion of
RIAFs and thus to reduce the outflow rate \citep[e.g.,][]{inayoshi2018low}.
This would yield a flatter radial profile of the mass inflow rate, consistent with the results in K21.
Finally, it is worth noting that in K21 the dense disk region is surrounded by a hot optically thin atmosphere set 
in the initial configuration, while dense optically thick outflows fill the polar regions in our cases.
We speculate that the difference might be caused by the initial and outer-boundary conditions.
We leave further investigation about this to future work.

\vspace{2mm}
\subsection{Simulations with Radiative Diffusion} \label{subsubsec:Diffusion}

Finally, we discuss the effect of radiative diffusion.
Note that the simulation setup with radiative diffusion (Case-D) is the same as the fiducial case (Case-H)
except that a lower resolution is adopted ($N_r \times N_\theta = 128\times 128$; see Table~\ref{tab:setups})
to avoid a significantly shorter diffusion timescale, i.e., $t_{\rm diff} \ll t_{\rm dyn}$.
For comparison, we also show simulation results of the fiducial model with the same lower grid resolution (Case-H-LR).

\begin{figure}[t!]
\centering
\includegraphics[scale=0.25]{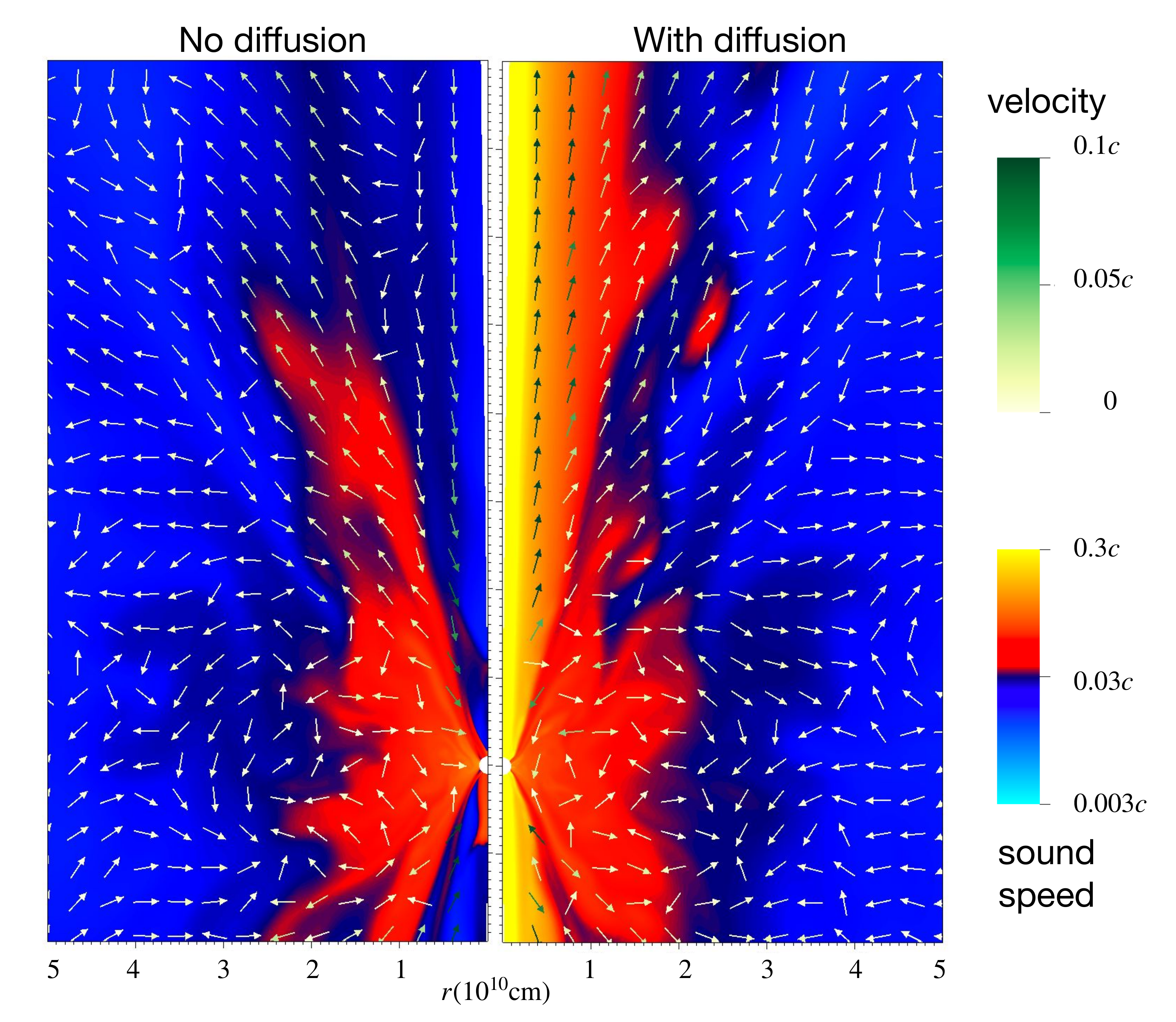}
\caption{2D distribution of the sound speed for Case-H-LR (left) and Case-D (right) when the system reaches a 
quasi-steady state ($t=2.4\times 10^6~t_0$). The velocity vectors are overlaid. 
The sound speed becomes higher in the polar region for Case-D owing to energy transport from the disk to the polar 
region. Radiative diffusion also plays an important role in driving the bipolar outflows at small scales, as indicated 
by the velocity vectors in the right panel. }
\label{fig:2d_cs}
\end{figure}

Fig.~\ref{fig:acc_diffusion} presents the time evolution of the mass flow rates (left panel)
and the time-averaged radial profiles (right panel) of these rates for the diffusion case (Case-D; dashed curves) and Case-H-LR (solid curves).
The red and blue curves correspond to the net accretion rate and the mass inflow rate, respectively.
Overall, the mass flow rates in the two cases are similar, although the net accretion rate increases slightly due to energy transport by 
radiative diffusion.
This suggests that energy transport via radiative diffusion plays a minor role in determining the bulk properties of 
highly optically thick accretion flows at the vicinity of the BH ($r<10^3~r_{\rm Sch}$),
where energy advection owing to photon trapping (the angle-averaged trapping radius for these two cases are $r_{{\rm tr, eff}}\simeq 50~r_{\rm Sch}$, 
inferred from rest-frame luminosity) dominates as expected in previous numerical and analytical studies.

The left panel of Fig.~\ref{fig:theta_profiles} presents the angular dependence of the density at two radii 
($r=10$ and $1000~r_{\rm Sch}$) for Case-H-LR (solid) and Case-D (dashed).
The profiles in the two cases are similar near the equator, but the density near polar regions at $r\simeq 10~r_{\rm Sch}$
in Case-D is an order of magnitude lower than that in Case-H-LR.
The low-density regions form due to mass loading into bipolar outflows accelerated by the strong pressure gradient force
due to radiation energy leakage from the photospheric surface. As a reference, we overlay the analytical result for the density distribution of a CDAF solution with 
$\gamma =4/3$ \citep[solid black curve;][]{Quataert2000}.
In both cases, the level of density concentration near the equator is different from 
the self-similar solution predicted by the CDAF solution. 
That is attributable to the fact that the radial density profile is as steep as $\rho\propto r^{-1}$ at large radii and energy leakage due to radiative diffusion at small radii.

\begin{figure}[t!]
\centering
\includegraphics[scale=0.55]{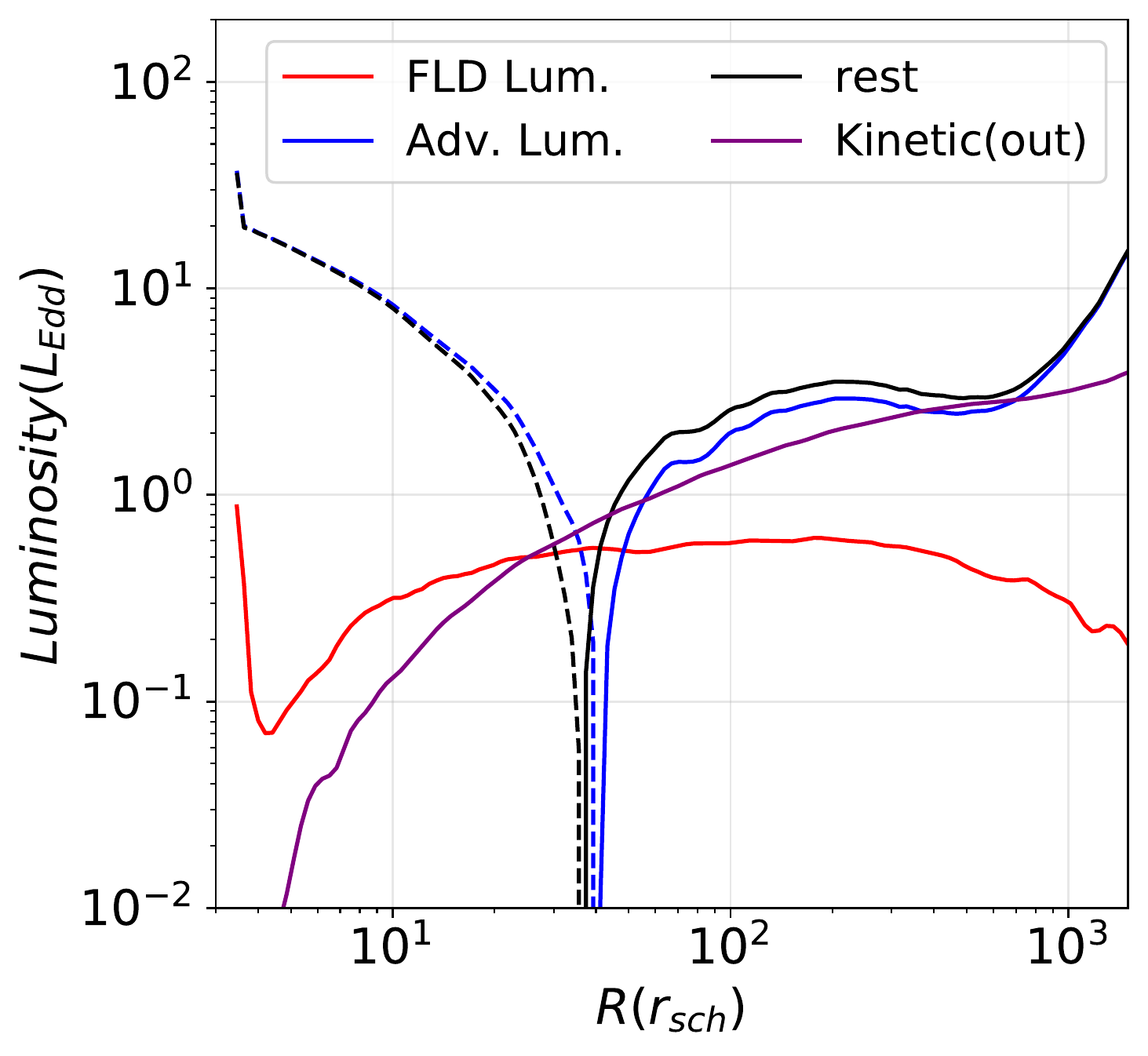}
\caption{Time-averaged radial profiles of three luminosities for the case with radiative diffusion (Case-D): 
the radiation luminosity in the rest (observed) frame (black), the advection luminosity (blue), the radiation luminosity 
in the fluid (comoving) frame (red), and the kinetic luminosity of outflows with $v_r > 0$ (purple). 
The rest-frame luminosity overall follows the advection component and thus it becomes negative 
due to inward advection (i.e., photon trapping) within $r \la 40~r_{\mathrm{Sch}}$.
}
\label{fig:flux_fid}
\end{figure}

The right panel of Fig.~\ref{fig:theta_profiles} shows the radial profiles of the density along the equator (red) 
and the polar angle $\theta = 30^\circ$ (blue) for the two cases.
Along the equator, the profiles are consistent with each other because radiative diffusion does not play an important role
in the highly optically thick dense region.
On the other hand, the density profile near the polar region at $r\simeq 10~r_{\rm Sch}$ is affected by radiative diffusion 
because a larger fraction of radiation energy is transported via diffusion into the polar region.
The radiative diffusion effect in mildly optically thick layers creates a strong pressure gradient, which pushes the gas 
outward, and thus the outflowing gas piles up near $\sim 30~r_{\rm Sch}$. 
Nevertheless, the overall impact by radiative diffusion is limited near the inner and polar regions.

\begin{figure}
\begin{center}
\includegraphics[scale=0.22]{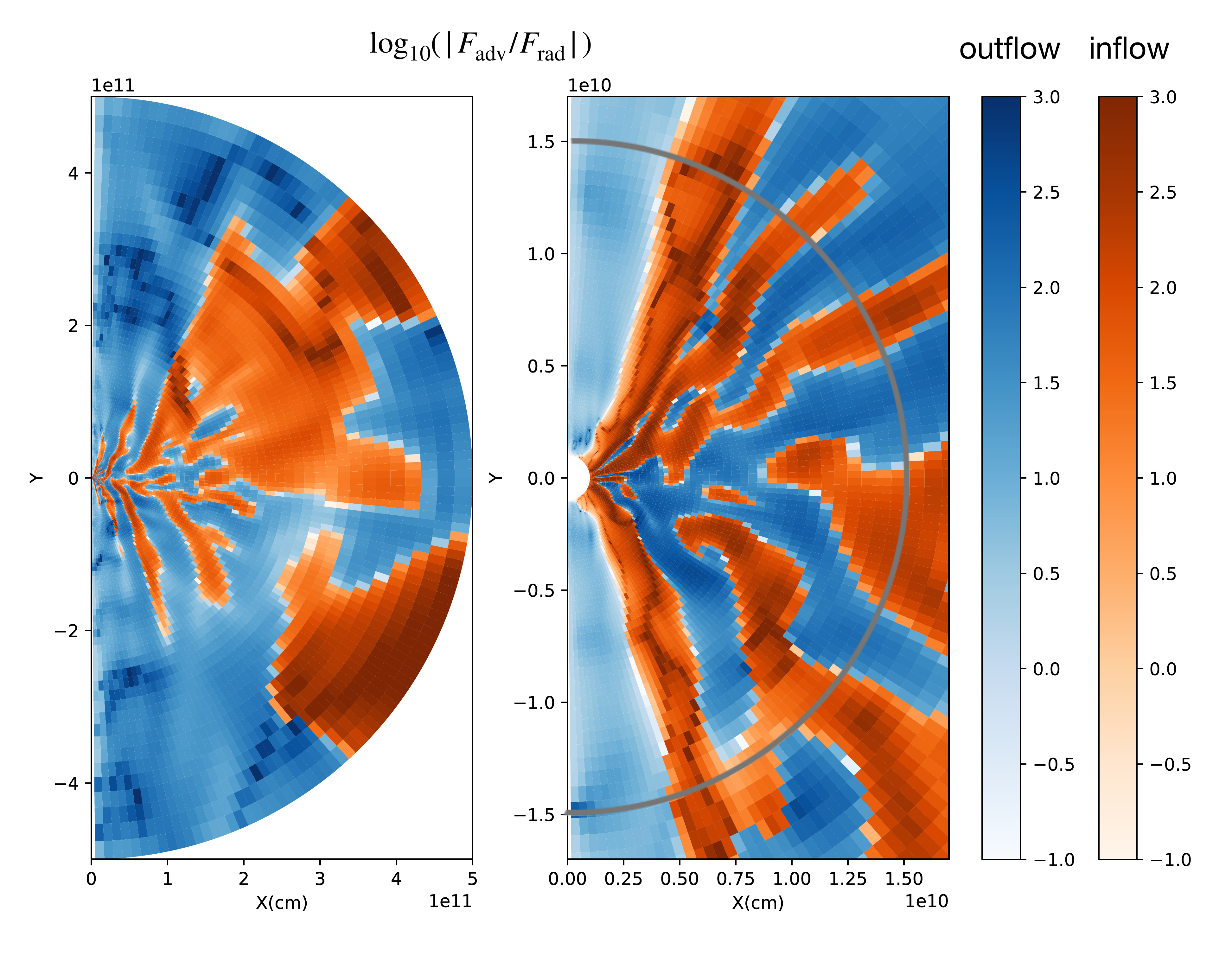}
\caption{Two-dimensional distribution of the ratio of the advection flux (blue for outflow, 
orange for inflows) to the radiative flux ($\log_{10}(|F_{\rm adv}/F_{\rm rad}|)$) in the comoving frame at 
$t=3.0\times 10^6~t_0$ for Case-D. 
The two panels show the distribution in the whole region (left) and in the inner region within the effective trapping radius (right). 
The effective trapping radius inferred from the rest 
frame luminosity is overlaid ($\sim 40~r_{\rm Sch}$, the grey half circle in the right panel). For reference, the spherical photon-trapping radius is $r_{\rm tr}\sim 6000~r_{\rm Sch}$.
The radiative flux dominates in the polar regions and the advective flux dominate in the equatorial regions. 
The 2D distribution does not show a distinct spherical surface characterizing the photon trapping radius.}
\label{fig:adv}
\end{center}
\end{figure}

In Fig.~\ref{fig:2d_cs}, we present the 2D distribution of the sound speed for Case-H-LR (left) and Case-D (right) when the systems reach a 
quasi-steady state ($t=2.4\times 10^6~t_0$). This clearly shows that the radiation energy is mainly transported from the 
disk and increases the sound speed in the polar regions significantly. We also find that the bipolar outflows are accelerated 
from the vicinity of the BH horizon owing to strong pressure gradient and lead to a steeper density gradient (see also 
Fig.~\ref{fig:theta_profiles}). Note that although the outflow velocity in the diffusion case reaches $\gtrsim 0.1~c$ 
\footnote{The maximum outflow speed reaches at most $\simeq 0.3~c$, which corresponds to a Lorentz factor of 1.05. 
Thus, the relativistic effects do not change our simulation results significantly.}, this effect increases the outflow rate by a 
factor of $\sim 1.3$ within the trapping radius compared to the case without radiative diffusion (see the right panel of Fig.~\ref{fig:acc_diffusion}).
The faster outflow at the vicinity of the launching scale does not increase the outflow momentum output at larger radii
because the acceleration mechanism owing to radiative diffusion operates only at the small radii.

For the highly accreting gas at rates of $\dot{M}_{\rm in} \gg \dot{M}_{\rm Edd}$, the radiative and mechanical outputs onto 
larger scales are crucially important to decide the efficiency of feedback, which in turn governs the gas supply rate 
onto the nuclear region of interest.
In a super-Eddington accreting system, photon trapping is considered to be an important effect to reduce the radiative output 
and thus moderate the feedback strength.
With a steady, spherically symmetric flow, the size of the trapping radius is estimated as 
$r_{\rm tr}/r_{\rm Sch} \simeq 5000~(\dot{M}_{\rm in}/10^3~\dot{M}_{\rm Edd})$.
However, this simplest assumption is broken in our axisymmetric 2D simulations owing to strong bipolar (non-spherical) outflows, 
which reduce the mass inflow rate toward the center.
Therefore, we need to quantify the nature of photon trapping via relative contributions of different energy transport mechanisms.

\begin{figure*}
\centering
\includegraphics[scale=0.4]{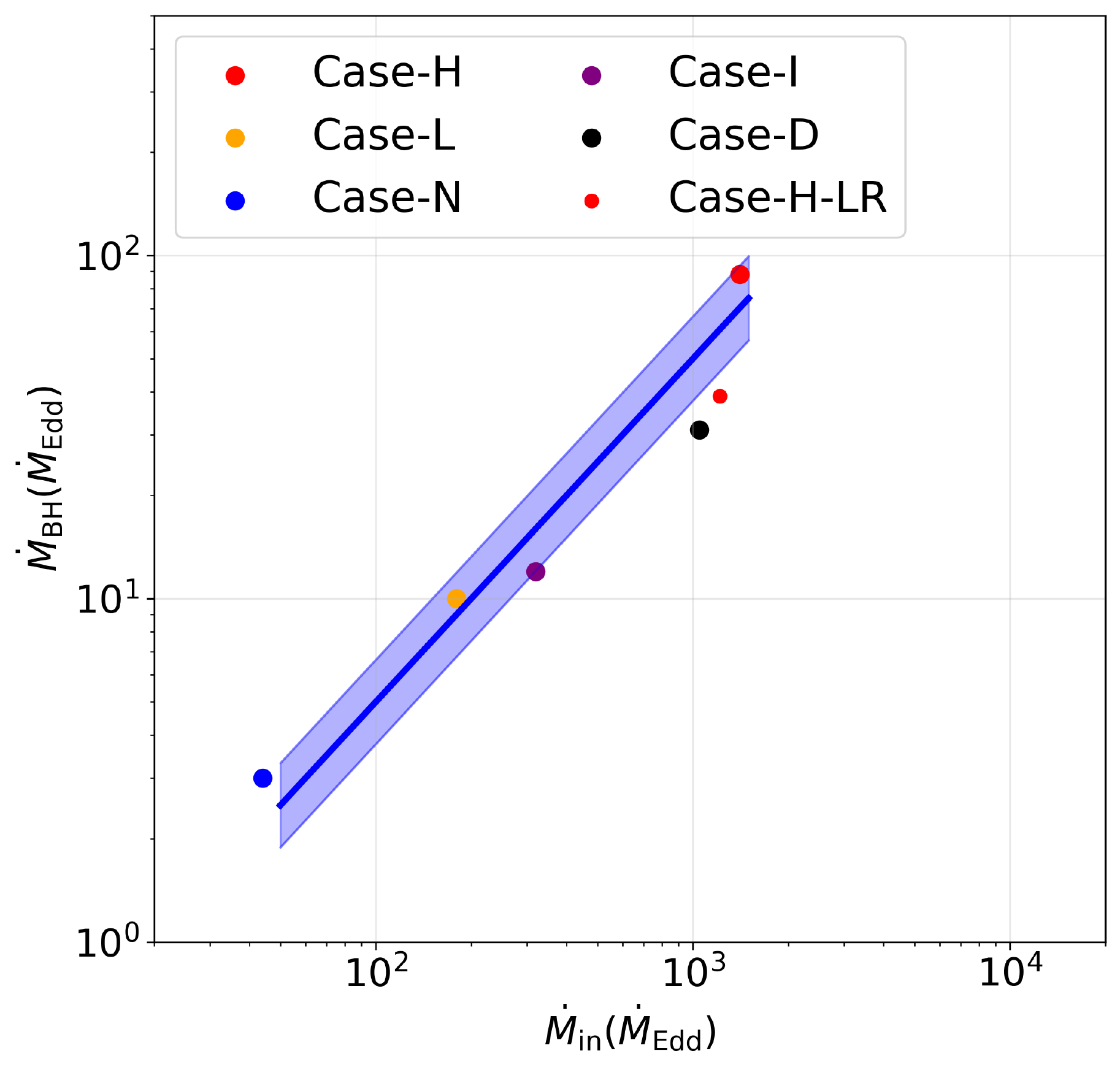}\hspace{10mm}
\includegraphics[scale=0.4]{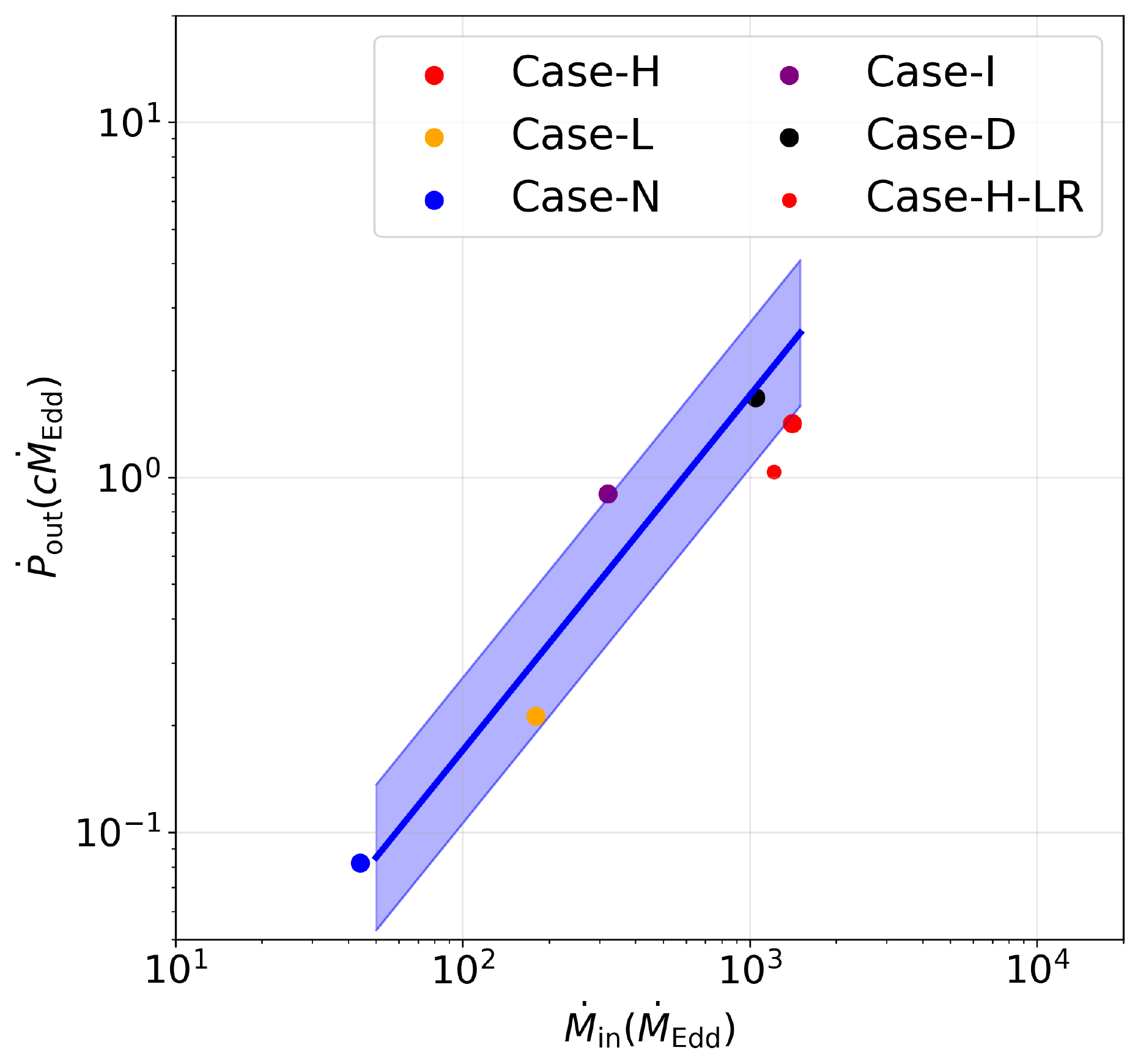}
\caption{The relations between the inflow at the outer boundary ($\dot{M}_0\equiv \dot{M}_{\rm in}(r=r_{\rm max})$) 
and the net BH accretion rate (left panel) and the 
momentum ejection rate of the outflow (right panel) in our six simulations with different inflow rates. The solid lines in 
both panels present the fitted linear relations given in Eqs.~(\ref{eq:Mdotout}) and (\ref{eq:Pdotout}). 
Each quantity is time-averaged and normalized by its Eddington value. 
Note that the results with a lower resolution for Case-D and Case-H-LR yield BH feeding rates nearly half of the high 
resolution case (Case-H). This small but systematic offset is comparable to those for other cases around the linear 
relation and thus does not affect our conclusion. The blue-shaded regions mark the 1 $\sigma$ deviation relative to the scaling relations.
We also note that despite radiative diffusion effects in accelerating the outflow in the inner region near the poles (see Fig.~\ref{fig:2d_cs}), 
the outflow momentum flux at outer boundary for Case-D (compared with Case-H-LR) is not significantly affected due to the fact that 
radiative diffusion has no effects on large scale.
}
\label{fig:acc_relation1}
\end{figure*}

In Fig.~\ref{fig:flux_fid}, we show the radial profiles of four kinds of luminosities for Case-D: 
the radiation luminosity in the rest (observed) frame (black),
\begin{equation}
L_{\mathrm{rest}} =L_{\mathrm{com}} + L_{\mathrm{adv}},
\label{eq:rest}
\end{equation}
 the advection luminosity (blue), 
 \begin{equation}
L_{\mathrm{adv}} = \int_{0}^{\pi}2\pi r^2   (4v_r p)\sin\theta d\theta,
\label{eq:adv}
\end{equation}
the radiation luminosity in the fluid (comoving) frame (red),
\begin{equation}
L_{\mathrm{com}} = \int_{0}^{\pi}2\pi r^2 {F}_{0,r} \sin\theta d\theta,
\label{eq:FLD}
\end{equation}
 and the kinetic luminosity of outflows with $v_{r}>0$ (purple)
 \begin{equation}
L_{\mathrm{kin,out}} = \int_{0}^{\pi}2\pi r^2   \left(\frac{1}{2}\rho v^2 \cdot \max(v_r,0)\right)\sin\theta d\theta.
\label{eq:kin+}
\end{equation}

Note that if the gas were static, i.e. $v_ r=0$, the comoving luminosity would be identical to the rest-frame luminosity.
In the optically thick flow, the comoving luminosity calculated by the FLD method is almost constant at $\sim 0.6~L_{\mathrm{Edd}}$
over the simulation domain.
This value is consistent with the radiation luminosity that diffuses out in a spherically symmetric flow \citep{begelman1979can}.
On the other hand, the advection luminosity becomes negative (dotted) in the inner region of $r \la 40~r_{\mathrm{Sch}}$, 
while it becomes positive (solid) at larger radii owing to strong outflows.
The rest-frame luminosity basically follows the advection component and thus it becomes negative 
within $r \la 40~r_{\mathrm{Sch}}$, which we can regard as the effective photon-trapping radius. 
The radiation-dominated outflow also carries kinetic energy and the kinetic luminosity reaches $\sim 3~L_{\rm Edd}$, 
which is consistent with that obtained by an analytical model for an optically thick outflow \citep[see][]{Piran2015}.

The effective photon-trapping radius defined in Fig.~\ref{fig:flux_fid} is an angle-averaged value. 
As an alternative way to demonstrate the photon trapping effect, Fig.~\ref{fig:adv} presents the 2D distribution 
of the ratio of the advective flux for the inflow component to the comoving radiative flux at $t=3.0\times 10^6~t_0$ 
on the largest scale (left) and in the inner region of $r \la 40~r_{\rm Sch}$ (the outflow regions are shown by blue). 
Overall, the radiative flux dominates the advective flux for outflows near the polar region, as indicated by light blue regions in the right 
panel. The advection flux dominates in dense and optically thick regions, as indicated by regions with dark blue and orange colors 
(right panel in Fig.~\ref{fig:theta_profiles} and equatorial region in Fig.~\ref{fig:adv}). 
Importantly, the 2D distribution does not show a distinct characteristic spherical surface, within which 
the advection flux overcomes the diffusive flux. It is reasonable that the angle-averaged (effective) trapping radius ($30-50~r_{\rm Sch}$) is 
much smaller than the spherical trapping radius ($220-7000~r_{\rm Sch}$). As shown in Table~\ref{tab:setups}, the effective trapping radius is 
always near $r_{{\rm tr, eff}}\sim 30-50~r_{\rm Sch}$ with a weak dependence on the gas supply rate.

\section{Subgrid Models for Cosmological Simulations}\label{sec:subgrmodel}

In general, the nature of BH feeding and feedback is determined by the global response of the accreting and outflowing gas
over a wide range of scales, from the vicinity of BHs ($r_{\rm max} \sim 10^{-7}$ pc for $M_{\rm BH}=10^3~M_{\odot}$) to 
galactic scales ($\sim 1$ kpc). 
The required dynamical range of the computational domain is therefore about 10 orders of magnitude in radius, 
which makes such simulations computationally challenging. 
Recently, \citet{Lalakos2022} have conducted MHD simulations of radiatively inefficient  accretion flows 
(i.e., substantially low accretion rates) that cover 5 orders of magnitude in space to study the jet launching mechanism 
associated with MHD effects and its impact on the gas at larger scales. 
Similar types of RHD simulations in a sufficiently long term \footnote{\citet{Lalakos2022} follows the evolution of the 
accretion system to $2\times10^5~t_0$, which is substantially shorter than the viscous timescale at $1000~r_{\rm Sch}$.} 
are required to better understand the global interaction between super-Eddington accretion flows and outflows. 
An alternative approach for this problem is to separate the computation domain into several regions and to treat
the dynamics of accretion and feedback crossing those regions in a self-consistent way. For example, \citet{Botella2022} showcase the impact of outflows originating from supercritical accretion, using nested large-scale simulations. However, they limit their study to a specific configuration, and do not provide prescriptions that  can be used in other large scale simulations. 
For this purpose, we provide a subgrid feedback model for super-Eddington accretion flows accompanied by 
strong outflows that carry mass and momentum.

\begin{figure}
	\centering
	\includegraphics[scale=0.32]{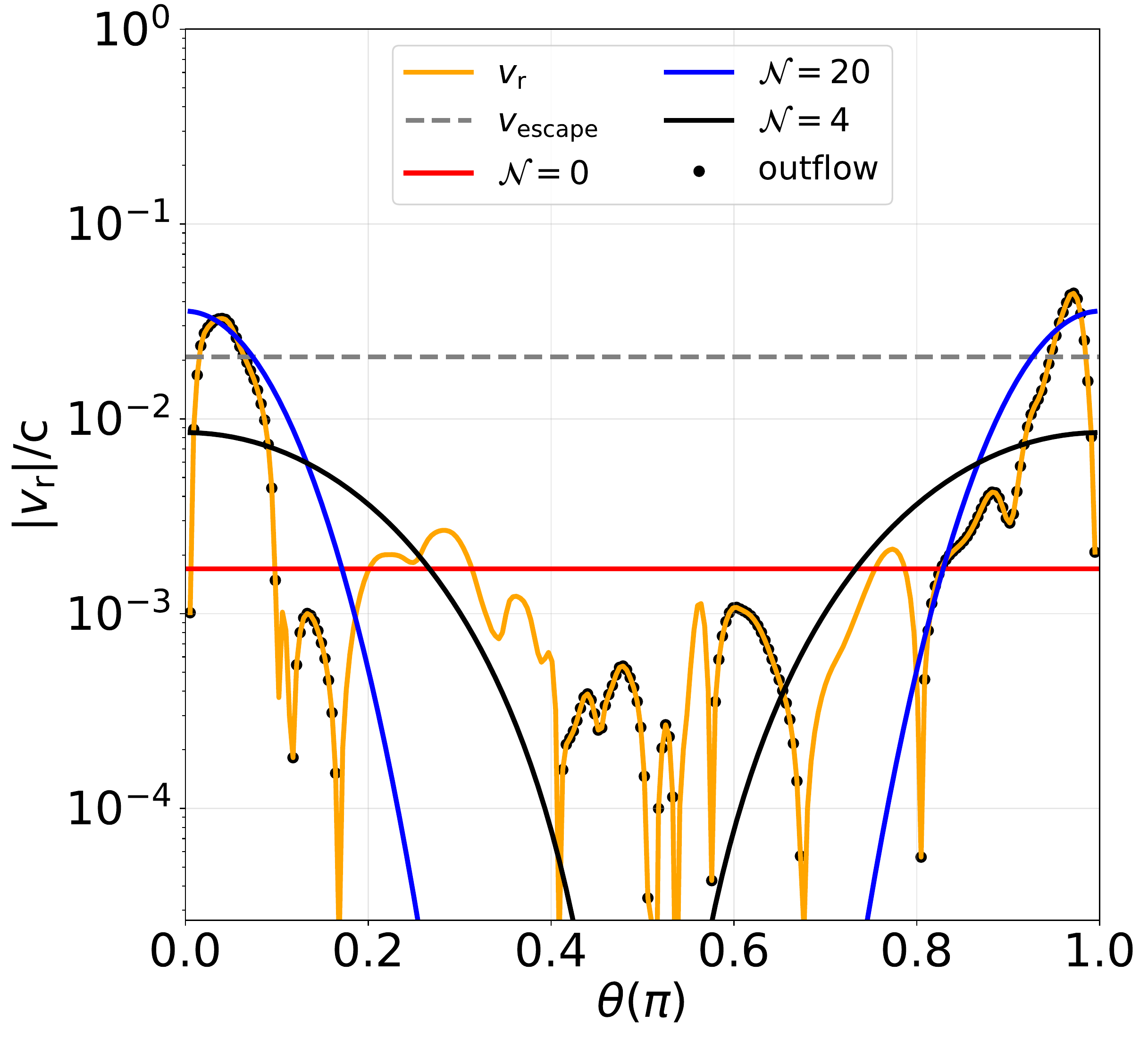}
	\caption{ The angular distribution of the time-averaged radial velocity at the outer-most radius (orange curve) 
	with black dots denoting the outflow components $(v_r>0)$. 
	The outflow velocity in the analytical expression with Eqs. (19) and (20) is shown; $\mathcal{N}=0$ (red), 
	$\mathcal{N}=4$ (black), and $\mathcal{N}=20$ (blue), where $\mathcal{N}$ characterizes the degree of anisotropy
	 of the outflow. The case with $\mathcal{N}=20$ describes the simulation results near the poles well, with the velocity 
	 reaching the escape velocity at $r=r_{\rm max}$ (horizontal gray line).}
	\label{fig:vr_angular}
\end{figure}

In Fig.~\ref{fig:acc_relation1}, we present the BH accretion rate (left) and the outflow momentum flux (right) 
as a function of the inflow rate (each point corresponds to the case denoted in the figure).
These three quantities show nearly linear correlations as 
\footnote{Without imposing the linear relation, the best-fit 
slopes are slightly different from unity; namely $\dot{M}_{\rm BH} \propto \dot{M}_0^{0.95}$ and $\dot{P}_{\rm out} \propto 
\dot{M}_0^{0.86}$. Note that Case-D and Case-H-LR are not taken into account for the fitting data, because the two cases 
are lower-resolution runs and yield nearly twice lower BH accretion rates. We here assume that the power-law distribution 
of the mass inflow rate ($\dot{M} \propto r^p$) is maintained until the exterior of the computational domain 
(presumably, until the spherical photon-trapping radius).}
\begin{equation}
\dot{M}_{\rm BH}=5.7^{+1.8}_{-1.3}\times 10^{-2}~\dot{M}_0 \left(\frac{r_{\rm max}}{1500~r_{\rm Sch}}\right)^{-0.5},
\label{eq:Mdotout}
\end{equation}
\begin{equation}
\dot{P}_{\rm out}= 1.7^{+0.9}_{-0.6}\times 10^{-3}~c\dot{M}_0
\left(\frac{r_{\rm max}}{1500~r_{\rm Sch}}\right)^{-0.5},
\label{eq:Pdotout}
\end{equation}
where $\dot{M}_0\equiv \dot{M}_{\rm in}(r=r_{\rm max})$ and $p=0.5$ is adopted. 
For reference, the values of $\dot{M}_0$ for the six cases with different outer boundary conditions are summarized in Table~\ref{tab:setups}.
From Eqs.~(\ref{eq:Mdotout}) and (\ref{eq:Pdotout}), one can obtain the mass loading factor and 
the angular-averaged radial velocity of the outflow $\langle v_{\rm wind}\rangle (\equiv \dot{P}_{\rm out}/\dot{M}_{\rm out})$,
respectively, as
\begin{equation}
\beta =  17.5^{+5.5}_{-4.2}~\left(\frac{r_{\rm max}}{1500~r_{\rm Sch}}\right)^{0.5} -1,
\end{equation}
\begin{equation}
\langle v_{\rm wind}\rangle = 1.7^{+0.9}_{-0.6}\times 10^{-3}c ~\left(\frac{r_{\rm max}}{1500~r_{\rm Sch}}\right)^{-0.5}.
\end{equation}
It is worth noting that the outflow is collimated to the polar regions and the outflow velocity is higher than the escape velocity 
within the bi-conical regions. 
Here, we model the angular profile of the outflow velocity as
\begin{equation}
v_{\rm wind} = \langle v_{\rm wind}\rangle (\mathcal{N}+1) \cos^{\mathcal{N}} \theta.
\label{eq:v_wind}
\end{equation}
where $\mathcal{N}$ is a free parameter that characterizes the anisotropy of the outflow, and is 
determined so that the outflow velocity near the poles exceeds the escape velocity. 
Fig.~\ref{fig:vr_angular} presents the radial velocity in Case-H 
(orange curve; the dots represent the outflowing component), the angle-averaged wind velocity (red curve), 
and the models from Eq.~(\ref{eq:v_wind}). 
This figure suggests that the case with $\mathcal{N}=20$ agrees well with the simulation results near the poles and 
the outflow velocity near the pole exceeds the escape velocity (grey line).

We note that the relations shown above are valid when the mass inflow rate through the outer-most radius 
is substantially higher than the Eddington value so that the spherical photon-trapping radius for the given $\dot{M}_0$ 
is larger than the simulation domain.  
Therefore, if the spherical photon-trapping radius is resolved (though this is not the case in our simulations), 
the outer-most radius would be replaced with the trapping radius ($r_{\rm tr} = 5\dot{m}_0~r_{\rm Sch}$,
where $\dot{m}_0 \equiv \dot{M}_0/\dot{M}_{\rm Edd}$).
Substituting this $r_{\rm tr}$ for $r_{\rm max}$, we can rewrite the above four equations as
\begin{equation}
\dot{M}_{\rm BH}=17.1~\dot{M}_{\rm Edd} \left(\frac{\dot{m_0}}{300}\right)^{0.5},
\label{eq:1}
\end{equation}
\begin{equation}
\dot{P}_{\rm out}= 0.51~c\dot{M}_{\rm Edd}
\left(\frac{\dot{m_0}}{300}\right)^{0.5},
\label{eq:2}
\end{equation}
\begin{equation}
\beta =  17.5~\left(\frac{\dot{m_0}}{300}\right)^{0.5} -1,
\label{eq:3}
\end{equation}
and
\begin{equation}
\langle v_{\rm wind}\rangle = 1.7\times 10^{-3}c ~\left(\frac{\dot{m_0}}{300}\right)^{-0.5}.
\label{eq:4}
\end{equation}
Note that those quantities are supposed to be the values at larger distances outside the trapping radius, 
where no further production and acceleration of the outflow is considered. Therefore, 
assuming that the mass accretion rate through sink cells (or onto sink particles)
is given by $\dot{M}_0$, we propose to apply Eqs.~(\ref{eq:v_wind})-(\ref{eq:4}) to large-scale cosmological simulations 
that hardly resolve the nuclear region but take into account mechanical feedback accompanied by super-Eddington
accreting BHs ($\dot{m}_0\gg 10$).

As an important caveat, our simulations focus on axisymmetric accretion flows without taking into account MHD effects explicitly, 
but adopting the $\alpha$-viscosity prescription ($\alpha = 0.01$) to treat angular momentum transport.
With a toroidal magnetic field or multiple poloidal loops configuration in the initial conditions or in injected gas, 
the accretion flow is dominated by turbulent motion driven by MRI and the saturated magnetic energy density is limited to
less than $0.1-1\%$ of thermal energy density. As a result, this type of accretion flow is qualitatively similar to that 
obtained from non-MHD simulations.
In contrast, when a poloidal magnetic field configuration is assumed in an accretion disk around a rapidly spinning BH, 
strong outflows and/or jets tend to be launched owing to amplified magnetic fields near the BH \citep{Tchekhovskoy2011,McKinney2014,skadowski2015global,Tchekhovskoy2016},
which may further suppress the BH feeding rate. 
We leave a study of strongly magnetized outflows from super-Eddington accretion to a future investigation.

\section{Summary} \label{sec:summary}

In this paper, we study the long-term evolution of the global structure of axisymmetric accretion flows onto a black hole fed on large scales at rates 
substantially higher than the Eddington value, performing 2D RHD simulations. 
The global structure of the flow settles down to a quasi-steady state in millions of the orbital timescale at the BH event horizon, 
which is $\ga 10-100$ times longer than that addressed in previous (magneto-) RHD simulation studies.

In the high-accretion optically thick limit, where the radiation energy is efficiently trapped within the flow,
the accretion flow becomes adiabatic and thus comprises of turbulent gas in the equatorial region and strong bipolar outflows. 
As a result, the mass inflow rate decreases toward the center as $\dot{M}_{\rm in}\propto r^{p}$ with $p\sim 0.5-0.7$, 
and only a small fraction of the inflowing gas ends up accreted by BH (see Figs.~\ref{fig:mass_flow_fid}, \ref{fig:acc_all} and 
\ref{fig:acc_diffusion}). In our simulations, super-Eddington accretion is sustained only 
when a larger amount of gas is supplied from larger radii at $\ga 100-1000~\dot{M}_{\rm Edd}$. Despite different gas supply 
rates owing to different outer-boundary conditions, the bulk dynamical properties of the gas inflow in all cases we studied show a 
self-similarity (see Figs.~\ref{fig:acc_all} and \ref{fig:profiles_all}). Even in the super-Eddington accretion regime, 
a fraction of the radiation energy in the flow can escape by diffusing out. Energy transport via radiative diffusion accelerates 
the outflow near the poles in the inner region but does not affect the overall properties of the accretion flow.

As for the strength of the outflows, a small value of $p~( \leq0.5-0.7 )$ value is expected for supercritical accretion flows since 
most of the radiation is thought to be trapped near the BH. Simulations with different scales and initial setups show that the value of $p$ is near $0.5-1$, 
with the uncertainties coming from the treatment of radiation feedback and the strength of the gas supply. 
Further studies of the global properties of outflows and inflows are necessary to determine the outflow strength in a self-consistent manner.

Cosmological and galactic scale simulations cannot resolve the inner regions of the BH accretion flows 
and need to implement BH growth and feedback with sub-grid descriptions \citep[e.g.][]{Massonneau2022}.
In this paper, we provide simple formulae (see Eqs.~\ref{eq:v_wind}-\ref{eq:4}), which can be used in simulations
of super-Eddington accretion at their resolution limit, to predict the BH accretion rate, as well as
the mass loading factor, velocity, and momentum injection rate of the outflow, as functions of the mass inflow rate
at the resolution limit.
This physically motivated implementation of outflow models can help us understand the co-evolution
between SMBH and host galaxies in a self-consistent way, connecting large and small scale simulations.

\begin{acknowledgements}
We thank the anonymous referee for a careful reading of our manuscript and comments that helped improved this paper.
K.I. acknowledges support from the National Natural Science Foundation of China (12073003, 
11721303, 11991052, 11950410493), and the China Manned Space Project with NO. CMS-CSST-2021-A06. Z.H. acknowledges support 
from NSF grant AST-2006176. R.K. acknowledges financial support via the Heisenberg Research 
Grant funded by the German Research Foundation (DFG) under grant No.$\sim$KU 2849/9. Further, R.K. 
acknowledges financial support via the JSPS Invitational Fellowship for Research in Japan under 
the Fellowship ID S20156. The numerical simulations were performed with the Cray XC50 
at the Center for Computational Astrophysics (CfCA) of the National Astronomical Observatory of 
Japan and with the High-performance Computing Platform of Peking University.
\end{acknowledgements}



\end{CJK*}
\end{document}